\documentclass{aa}

\usepackage{graphicx}
\usepackage{txfonts}
\usepackage{natbib}
\usepackage{longtable}
\bibpunct{(}{)}{;}{a}{}{,} % to follow the A&A style

\usepackage[shortlabels]{enumitem}

\usepackage{multicol}
\usepackage{multirow}

\usepackage[normalem]{ulem}
\usepackage{bm}
\usepackage{xspace}
\usepackage{xcolor}
\newcommand{\masyr}{mas\,yr\ensuremath{^{-1}}\xspace}%
\newcommand{\lsun}{\ensuremath{L_{\sun}}\xspace}

\newcommand{\micron}{\ensuremath{\mu}m\xspace}

\newcommand{\re}{\ensuremath{R_\textrm{e}}\xspace}
\newcommand{\mh}{[M/H]\xspace}
\newcommand{\kms}{km\,s\ensuremath{^{-1}}\xspace}
\newcommand{\msun}{\ensuremath{M_{\sun}}\xspace}
\newcommand{\mbh}{\ensuremath{M_{\bullet}}\xspace}
\newcommand{\col}{\ensuremath{H-K_{S}}\xspace}
\newcommand{\vlos}{\ensuremath{V_\textrm{LOS}}\xspace}

\newcommand{\slos}{\ensuremath{\sigma_{\textrm{LOS}}}\xspace}
\newcommand{\sgra}{Sgr~A$^\star$\xspace}

\begin{document}

  \title{Dynamical mass distribution and velocity structure of the Galactic centre} 
% \titlerunning{Dynamical mass distribution of the Galactic centre} 
\author{A. Feldmeier-Krause 
 \inst{1} \and
 T. Ver\v{s}i\v{c}\inst{1} 
  \and
 G. van de Ven\inst{1}
   \and
 E. Gallego-Cano\inst{2}
 \and
  N. Neumayer\inst{3}
}
 \institute{Department of Astrophysics, University of Vienna, T\"urkenschanzstrasse 17, 1180 Wien, Austria\\
 \email{anja.krause@univie.ac.at}
 \and
 Instituto de Astrofísica de Andalucía (CSIC), Glorieta de la Astronomía s/n, 18008 Granada, Spain
 \and
 Max Planck Institute for Astronomy, K\"onigstuhl 17, D-69117 Heidelberg, Germany
}

  \date{Received March 18, 2025; accepted  May 23, 2025}

% \abstract{}{}{}{}{} 
% 5 {} token are mandatory
 
 \abstract
  {The inner $\sim$200 pc region of the Milky Way contains a nuclear stellar disc and a nuclear star cluster that are embedded in the larger Galactic bar. These stellar systems overlap spatially, which makes it challenging to separate stars that belong to the nuclear stellar systems, to deduce their internal dynamics, and to derive the central Galactic potential.}
  {Discrete stellar kinematics probe the mass distribution of a stellar system, and chemical tracers such as stellar metallicity can further separate multiple stellar populations that can have distinct kinematic properties. 
  We took advantage of the information provided by discrete stellar kinematics and the metallicity of stars in the Galactic centre using discrete chemo-dynamical modelling. }
  {We fitted axisymmetric Jeans models to discrete data of 4\,600 stars. We fitted the stars as either one population plus a background component or as two populations plus a background that represents the bar. In the one-population case, we tested the robustness of the inferred gravitational potential against a varying mass of the supermassive black hole, including dark matter, or a radially varying mass-to-light ratio.  }
  {We obtained robust results on the stellar dynamical fit with a single population and a background component. We obtained a supermassive black hole mass of (4.35$\pm 0.24) \times 10^6$\,\msun, and we find that a dark matter component adds no more than a few percent to the total enclosed mass of the nuclear star cluster. The radial variation in the mass-to-light ratio is also negligible.  We derived the enclosed mass profile of the inner $\sim$60\,pc of the Milky Way and found a lower mass than reported in the literature in the region of $\sim$5-30\,pc.
  In our two-population fit, we found a high-\mh population with a mild tangentially anisotropic velocity distribution and stronger rotational support than for the low-\mh population, which is radially anisotropic. The high-\mh population is dominant and contributes more than 90\% to the total stellar density. }
  {The properties of the high-\mh population are consistent with in situ formation after gas inflow from the Galactic disc via the bar. The distinct kinematic properties of the low-\mh population indicate a different origin.   }


  \maketitle

%%%%%%%%%%%%%%%%%%%%%%%%%%%%%%%%%%%%%%%%%%%%%%%%%%

%%%%%%%%%%%%%%%%% BODY OF PAPER %%%%%%%%%%%%%%%%%%

\section{Introduction}

The Galactic centre region and its gravitational potential have been of great interest for several decades. Before the orbits of the stars around the supermassive black hole \sgra could be measured, which ultimately constrained its mass with unmatched accuracy and precision \citep[e.g.][]{2022A&A...657L..12G}, other methods were applied. Extended stellar and gas kinematics were used to constrain the mass distribution, including the mass of \sgra, \mbh. 
Similar methods are still used today to derive the mass distributions of star clusters and external galaxies. 

The Galactic centre contains two rotating stellar structures: The nuclear star cluster (NSC), with an effective radius \re$\sim$5\,pc and a flattening (minor-to-major axis ratio) $q\sim$0.7 \citep{2014A&A...566A..47S,2016ApJ...821...44F,2020A&A...634A..71G}, and the surrounding nuclear stellar disc (NSD), with scale length of $\sim$90\,pc, a scale height of 28-45\,pc, and $q\sim$0.35 \citep{2002A&A...384..112L,2013ApJ...769L..28N,2014A&A...566A..47S,2020A&A...634A..71G,2022MNRAS.512.1857S}. The NSC dominates the stellar surface density in the projected inner $r\sim$7\,pc of the Milky Way \citep{2025A&A...696A.213F}, and the NSD becomes dominant farther out.

Stellar dynamical models of the NSC often made assumptions on its shape and velocity structure. Many studies have assumed a spherical mass distribution and an isotropic velocity distribution
\citep[e.g.][]{1989ApJ...338..824M,1996ApJ...472..153G,1996ApJ...456..194H,2001AJ....122..232C,2008A&A...492..419T,2009JKAS...42...17O,2016ApJ...821...44F}. Some studies relaxed the assumption and allowed anisotropy \citep[e.g.][]{2000MNRAS.317..348G,2009A&A...502...91S,2013ApJ...779L...6D,2019MNRAS.484.1166M} and also a flattening of the cluster \citep{2014A&A...570A...2F,2015MNRAS.447..948C,2017MNRAS.466.4040F}.
Many of these studies focussed on data in the inner 1\,pc, and some had stellar kinematics data out to a few parsec of the Milky Way. Their goal often was to constrain the mass of \sgra, which they were not always able to do, and many underestimated it. 
A few works focussed on regions on the scale of $\sim$100\,pc
\citep{1992A&A...259..118L,2004PASJ...56..261D,2020MNRAS.499....7S,2022MNRAS.512.1857S,2024MNRAS.530.2972S}, where the flatter NSD dominates the gravitational potential. 
The intermediate region (5--40\,pc) lacked extended data so far, and the gravitational potential was inferred via interpolation.

One limitation of many stellar dynamical models is that they require the user to spatially bin the data to a mean velocity and velocity dispersion. By binning data, however, we lose information because we have to combine a sufficiently large number of velocities to obtain a robust estimate of the mean velocity and velocity dispersion. Only some of the aforementioned works on the Galactic centre used discrete data, but it has become more common in recent years \citep[e.g.][]{2001AJ....122..232C, 2013ApJ...764..154D,2019MNRAS.484.1166M,2022MNRAS.512.1857S,2024MNRAS.530.2972S}. Another disadvantage when the data are binned is that we have to assume that the stars that are considered for binning all belong to the system that is being modelled. This may not be true because the samples can be contaminated with foreground or background stars. Using discrete dynamical models, we can include such a contaminant component and assign each star a probability of being a member star or a contaminant \citep{2013MNRAS.436.2598W}.

It is also possible to include knowledge of the stellar populations in discrete models. For example, the stellar metallicity \mh 
can be used to differentiate multiple stellar populations, and their velocity structures can then be constrained separately. This has been done for Galactic globular clusters or dwarf spheroidal galaxies \citep{2016MNRAS.462.4001Z,2016MNRAS.463.1117Z,2020MNRAS.492..966K,2022ApJ...939..118K}. This approach is interesting for the Galactic centre because the NSC and NSD have a broad \mh distribution, and this indicates a mixture of stellar populations \citep{2015ApJ...809..143D,2017MNRAS.464..194F,2020MNRAS.494..396F,2021A&A...649A..83F}.
Most stars are metal rich, but a low-metallicity tail also extends to sub-solar metallicity (\mh$\gtrsim$--1.5\,dex). Previous studies suggested that the spatial distribution and/or kinematic properties of the sub-solar \mh stars are different from those of the dominating super-solar \mh stars \citep{2020MNRAS.494..396F,2020ApJ...901L..28D,2021A&A...650A.191S,2025A&A...696A.213F}. While they appear to rotate as fast as or even faster than the metal-rich stars in the NSC and inner NSD (\citealt{2020ApJ...901L..28D,2025A&A...696A.213F}; longitude $l\lesssim$30\,pc), they rotate much slower at larger scales in the NSD ($l\lesssim$200\,pc] and resemble the motion of stars in the Galactic bulge \citep{2021A&A...650A.191S}. Some sub-solar \mh stars may be remnants from past infall events of star clusters or a dwarf galaxy \citep{2020ApJ...901L..29A}, or they trace close passages of globular clusters \citep{2023A&A...674A..70I}.

We combine recent developments of discrete chemo-dynamical axisymmetric Jeans models with 
spectroscopic data presented in recent works (\citealt{2025A&A...696A.213F}; Xu et al. in prep.). The data cover the outer part of the NSC and extend out to longitude $l$=33\,pc along the Galactic plane. These data add new information that has been missing in previous models. In addition, we use spectroscopic data from the literature to cover the inner NSC \citep{2017MNRAS.464..194F,2020MNRAS.494..396F,2022MNRAS.513.5920F} and match the data with proper motion catalogues \citep{2016ApJ...821...44F,2021MNRAS.500.3213L,2025MNRAS.536.3707S}.

This paper is organised as follows: We describe the data in Sect. \ref{sec:data} and the modelling procedure in Sect. \ref{sec:model}. We summarise the results for one-population Jeans models in Sect. \ref{sec:result} and for two-population models in Sect. \ref{sec:res2pop}. We discuss our result in Sect. \ref{sec:discussion} and conclude in Sect. \ref{sec:con}.

\section{Data set}
\label{sec:data}
Our dynamical models require a stellar surface density profile and stellar kinematic data. The stellar surface density profile is used to predict the part of the gravitational potential that is generated by the stars. In addition, we included contributions of the gravitational potential to the supermassive black hole \sgra, and we tested how important it is to include a contribution from dark matter. Based on the adopted total gravitational potential, the dynamical model predicts the stellar kinematics. The latter are compared to the stellar kinematic data, which thus constrain the total gravitational potential. 
For the stellar kinematics, we used a combination of line-of-sight velocities \vlos and proper motions. 
To convert the proper motions from \masyr into \kms, we assumed a galactocentric distance of 8.3\,kpc \citep{2022A&A...657L..12G} for all stars. Thus, 1\arcsec\ corresponds to 0.04\,pc.
In addition, our data set contained measurements of the overall metallicity \mh and photometric colour \col for each star.

\begin{figure*}
 \centering
 \includegraphics[width=18cm]{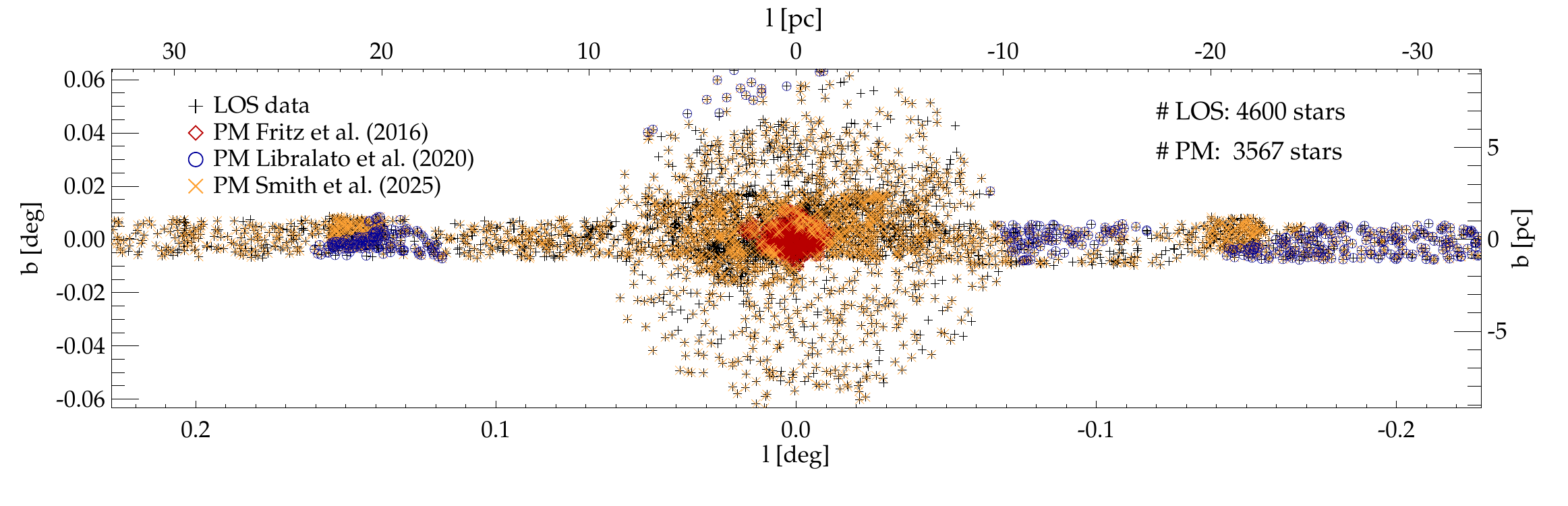}
  \caption{Spatial distribution of all stars we used for the model.
  The black crosses denote stars with \vlos\space and \mh, red diamonds show stars with additional proper motions from \cite{2016ApJ...821...44F},  blue circles show stars from \cite{2020MNRAS.497.4733L}, and orange crosses show stars from \cite{2025MNRAS.536.3707S}. }
  \label{fig:datapm}
\end{figure*}

\begin{figure}
\centering
 \includegraphics[width=0.9\columnwidth]{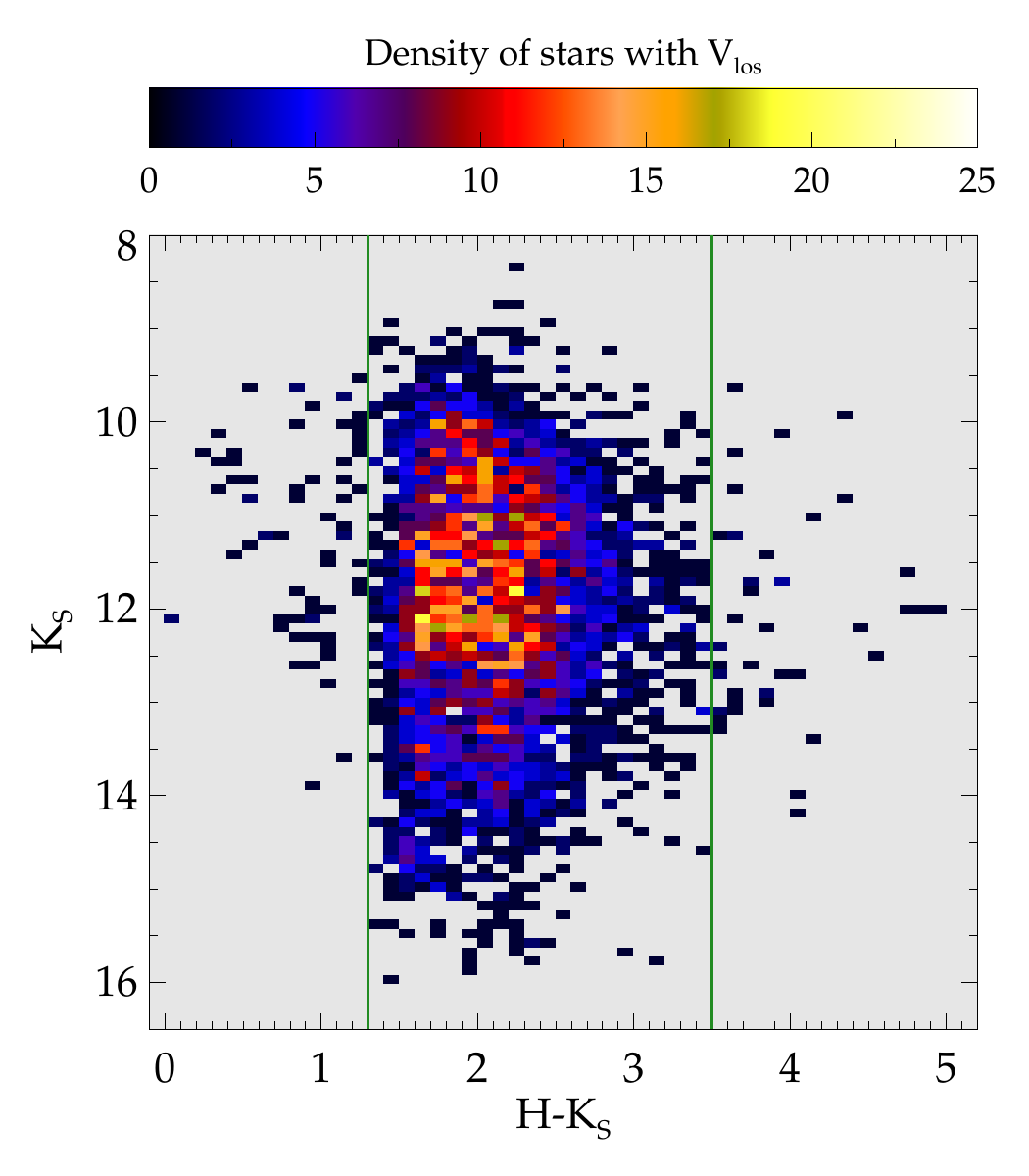}
  \caption{Colour-magnitude diagram $H-K_S$ vs. $K_S$ of the stars with \vlos\space and \mh. The vertical green lines denote the colour cuts we used to remove likely foreground (\col$\leq$ 1.3\,mag) and background stars (\col$\geq$3.5\,mag). The colour represents the density of stars.}
  \label{fig:cmddens}
\end{figure}

\subsection{Discrete line-of-sight velocity and metallicity catalogue}

We combined several data sets for the discrete stellar line-of-sight (LOS)  velocity \vlos\space and metallicity \mh. We show the spatial distribution of the combined catalogue in Fig.~\ref{fig:datapm} (black pluses). The density is highest in the centre because it contains the most stars. Density variations are also caused by the combination of different data sets that sample stars that are located in various regions and at varying depths.

\cite{2017MNRAS.464..194F} and \cite{2020MNRAS.494..396F} published measurements of \vlos\space and \mh\space of $\sim$700 stars each. The stars are located within a projected distance of 127\arcsec\ ($\sim$5\,pc) to Sgr~A*. \cite{2022MNRAS.513.5920F} published measurements for another $\sim$400 stars, half of which are located at a projected distance of $\sim$20\,pc (0.$^\circ$14) to the Galactic east and the other half lie west of \sgra. 
 Xu et al. (in prep.) provide a catalogue of $\sim$900 stars that are located within a circle with a radius of 247\arcsec ($\sim$0.$^\circ$067 $\sim$9.6\,pc) around \sgra. All these data were observed with KMOS\footnote{K-band multi-object spectrograph} \citep{2013Msngr.151...21S}  at the Very Large Telescope (VLT)  in the near-infrared $K$ band with a spectral resolution of $R$$\sim$4\,000. While the depth of the catalogues varies, the applied methods for deriving \vlos\space and \mh\space are identical between the various studies. In short, they used a full spectral fitting with the code \textsc{starkit} \citep{2015ApJ...809..143D,2015zndo.....28016K} and synthetic model spectra from the PHOENIX spectral library \citep{2013A&A...553A...6H}.

 For an even more extended data set, we used the catalogue of \cite{2025A&A...696A.213F}. These data were observed with Flamingos-2 at Gemini-South \citep{2004SPIE.5492.1196E}, also in the $K$ band, and at a slightly lower spectral resolution ($R$$\sim$3\,400). The data extend from \sgra\space to 33\,pc along the Galactic plane, both towards the Galactic east and west. The covered latitude is rather small, only $\sim$1\,pc to the Galactic south and 1\,pc (2\,pc in a central subfield) to the Galactic north. They applied the same method and models to derive \vlos\space and \mh.

 The data sets overlap to some extent. The Flamingos-2 data cover the same regions as the various KMOS catalogues, with an overlap of $\sim$60--300 stars, depending on the catalogue. We compared the \mh\space measurements and found that they are mostly consistent. The median \mh difference ($\Delta$\mh) of all stars with two \mh measurements is -0.09 to +0.04\,dex (depending on the KMOS data set; see also Appendix B of \citealt{2025A&A...696A.213F}). The median statistical uncertainty of all \mh is  0.18\,dex. The two \mh measurements are usually consistent within their  statistical uncertainties (i.e.  $\Delta$~[M/H]~/~$\sqrt(\sigma_1^2+\sigma_2^2) \lesssim $1,  where $\sigma_1$ and $\sigma_2$ denote the \mh uncertainties from the two independent measurements). 
For our final catalogue, we only combined the multiple measurements of a star when the catalogue coordinates agreed within 0\farcs2, the measurements of \vlos\space lay within 30\,\kms, and $K_S$ lay within 0.6\,mag. For these stars, we combined the multiple measurements of \vlos and \mh with a simple mean. As uncertainties, we used the propagated error of the mean (i.e. 0.5$\cdot \sqrt(\sigma_1^2+\sigma_2^2)$) or the standard deviation of the two measurements, whichever was larger. 
The \vlos\space was corrected for perspective rotation that is caused by the motion of the Sun around the Galactic centre. We computed this correction using the equations of \cite{2006A&A...445..513V}, and we found that it varied spatially by $\pm$1\,\kms at most.

Our final catalogue contains 4\,791 stars with \vlos\space and \mh\space measurements. All the stars are cool, with typical absorption lines that were used for the measurements of \vlos\ and \mh. The colour-magnitude diagram (CMD) in Fig.~\ref{fig:cmddens} shows the stars are on the red giant branch, but brighter than the red clump, which is at $K_S\sim$15\,mag in the Galactic centre. The stellar photometry comes from the GALACTICNUCLEUS (GNS) survey catalogue \citep{2019A&A...631A..20N}.

The CMD also shows that most stars have $H-K_S$\textgreater 1.3\,mag, indicating that they are near the Galactic centre. Stars with bluer colours are likely foreground stars. The catalogue contains only a few such stars, and we excluded stars with either $H-K_S$\textless 1.3\,mag or \col\textgreater 3.5\,mag as foreground and background stars, respectively. Our sample thus contains 4\,600 unique stars.

\subsection{Proper motion catalogues}
We complemented the  LOS data with the proper motion (PM) catalogues of \cite{2016ApJ...821...44F} for the inner $r\lesssim$89\arcsec ($\sim$3.6\,pc) and \cite{2020MNRAS.497.4733L} for outer regions $r\gtrsim$187\arcsec ($\sim$7.5\,pc).  The spatial distribution of the PM catalogues where they overlap with the LOS data is illustrated in Fig.~\ref{fig:datapm}. 
The proper motion catalogue of \citet[VIRAC\footnote{VVV Infrared Astrometric Catalogue} 2]{2025MNRAS.536.3707S} extends over the entire region of the LOS data, except for the most crowded inner $\sim$1\,pc region.  

We only used stars of the PM catalogues that also had entries in the LOS data to ensure that the stars were red giants and not hot young stars, because these last can be found across the Galactic centre \citep[e.g.][]{1999ApJ...510..747C,2010ApJ...710..706M,2010ApJ...725..188M,2011MNRAS.417..114D,2021A&A...649A..43C,2022MNRAS.513.5920F}.  
The spatial distribution \citep[e.g.][]{2013ApJ...764..154D,2015ApJ...808..106S,2015A&A...584A...2F, 2021A&A...649A..43C} and kinematics \citep[e.g.][]{2022ApJ...932L...6V,2022ApJ...939...68H} of hot young stars are different from those of the red giant stars we used as tracers of the gravitational potential. While the mass contribution of the young stars is negligible, their kinematics might still be used as independent tracers of the underlying total gravitational potential. The spatial distribution of young stars cannot be measured robustly, however, so that the resulting tracer density might bias the inferred gravitational potential. We therefore chose not to use young stars in our analysis.  The Galactic centre contains three young clusters, but only one of them  is located in the field of view of our data, in the central parsec. It has  a mass of $\sim$$10^4$\,\msun   \citep[e.g.][]{2006ApJ...643.1011P}.  The stellar surface density of young stars and red giants in the innermost parsec is comparable \citep{2013ApJ...764..154D}, but the gravitational potential in this region is dominated by \sgra. 
Farther out, the stellar number density of young stars decreases steeply, and they contribute only little overall to the mass at larger radii.

From the \cite{2025MNRAS.536.3707S} PM catalogue, we excluded stars with a PM exceeding 500\,\kms at a distance of 8.3\,kpc and stars with variable $K$-band photometry ($\sigma_{K_S}\geq$0.55\,mag) because their PMs are uncertain. Based on the resulting matches with the LOS catalogue (within 0\farcs2), we excluded stars with an $H$- or $K_S$-band discrepancy of more than 0.5\,mag as likely mismatches. This left 2\,688
stars with \cite{2025MNRAS.536.3707S} PMs. We found 875 matches of the LOS data with \cite{2016ApJ...821...44F} and 493 matches with \cite{2020MNRAS.497.4733L}. For the stars with PM measurements in two data sets ($\lesssim$500 stars), we computed the mean values. We obtained a PM measurement from at least one of these catalogues for 3\,567 unique stars. 

The PM catalogues list the PMs in equatorial coordinates, and we converted them into Galactic coordinates using the equations in \cite{2013arXiv1306.2945P}. 
The \cite{2020MNRAS.497.4733L} and \cite{2025MNRAS.536.3707S} data are in the absolute reference system of the Gaia DR2 catalogue. We corrected their PMs for the motion of \sgra\space relative to the Sun using the measurements of \cite{2020ApJ...892...39R} of -6.411\,\masyr\space along the Galactic plane and -0.219\,\masyr\space towards the Galactic north pole. A correction for perspective rotation, as made for the \vlos data, was unnecessary because the Sun does not move significantly towards or away from \sgra.

\subsection{Stellar density distribution and multi-Gaussian expansion fit}
\label{sec:mge}

We used the stellar density map of \cite{2020A&A...634A..71G}, produced from a combination of HAWK-I\footnote{High Acuity Wide field K-band Imager}/VLT \citep{2008A&A...491..941K} and NACO\footnote{Nasmyth Adaptive Optics System (NAOS) Near-Infrared Imager and Spectrograph (CONICA)}/VLT \citep{2003SPIE.4839..140R} $K_S$-band data. The HAWK-I data were obtained as part of the GNS survey \citep{2018A&A...610A..83N,2019A&A...631A..20N} and cover a region of 84.4\,pc$\times$21\,pc with a point spread function of 0\farcs2 full width at half maximum. The central region, where crowding is severe, was complemented with NACO data of $\sim$3.5\,pc$\times$3.5\,pc \citep[details on the data reduction in][]{2018A&A...609A..26G}. The NACO data have the benefit of a higher spatial resolution (0\farcs05) due to adaptive optics.

The data were cleaned from spectroscopically classified early-type stars (i.e. hot young OB-type stars) using the catalogue of \cite{2013ApJ...764..154D}, and the density map was mostly created from red giant stars, but it may contain a few early-type stars that were not yet spectroscopically classified. The fraction of early-type stars drops rapidly beyond the central $\sim$1\,pc region of the Galactic centre, however \citep{2015ApJ...808..106S, 2015A&A...584A...2F}. More extended spectroscopic studies, including those that provide our discrete catalogue, found only a few isolated early-type stars (\citealt{2022MNRAS.513.5920F,2025A&A...696A.213F}, Xu et al. in prep.). We conclude that the contamination of early-type stars in the density map is  $\lesssim$2\% and is therefore negligible.

The stellar photometry was corrected for extinction before  only stars in the range 9.0$\leq$$K_{S,\mathrm{ext}}$$\leq$14.0\,mag were  selected. This corresponds to 
a magnitude range $\sim$11.0$\leq$$K_S$$\leq$16.0\,mag. Hence, there is significant overlap with the kinematic tracer stars, which mainly lie at 9.0$\leq K_S \leq$15.0\,mag (see Fig.~\ref{fig:cmddens}). 
The photometric data are deeper and have a higher completeness than the kinematic tracer stars. 

Regions with high extinction can severely affect the density of detected sources, and \cite{2020A&A...634A..71G}  masked them in the stellar density map. 
The final map had a pixel scale of 5\,arcsec$\cdot$pixel$^{-1}$ and covered 84.4\,pc along the Galactic longitude and 21\,pc along the Galactic latitude. It is currently the largest and most complete stellar number density map of the Galactic centre.

The discrete Jeans models use the stellar surface brightness distribution of the tracer stars in the form of a multi-Gaussian expansion \citep[MGE,][]{1994A&A...285..723E}. We derived it using the \textsc{python} package \textsc{MgeFit} provided by \cite{2002MNRAS.333..400C}. 
We measured the counts of the density map in 5-degree-wide sectors centred on \sgra. 
We used our knowledge of the position angle of the major axis (along the Galactic plane), and the flattening (minor- to major-axis ratio) of the NSC \citep[$q \sim 0.71$,][]{2020A&A...634A..71G}. 
In the MGE fit, we constrained the value $q$ to \textgreater0.3 because the flattening of the NSD is larger than $\sim$0.3 \citep{2020A&A...634A..71G,2022MNRAS.512.1857S}. 
We further set the \textsc{outer\_slope} keyword, which forces the model to a profile at least as steep as $R^\mathrm{-outer\_slope}$, to 0.5, to allow for a flat profile at large radii. The default range of 1--4 in \textsc{MgeFit}  is too steep for the Galactic centre.

\begin{figure}
\centering
	 \includegraphics[width=0.9\columnwidth]     {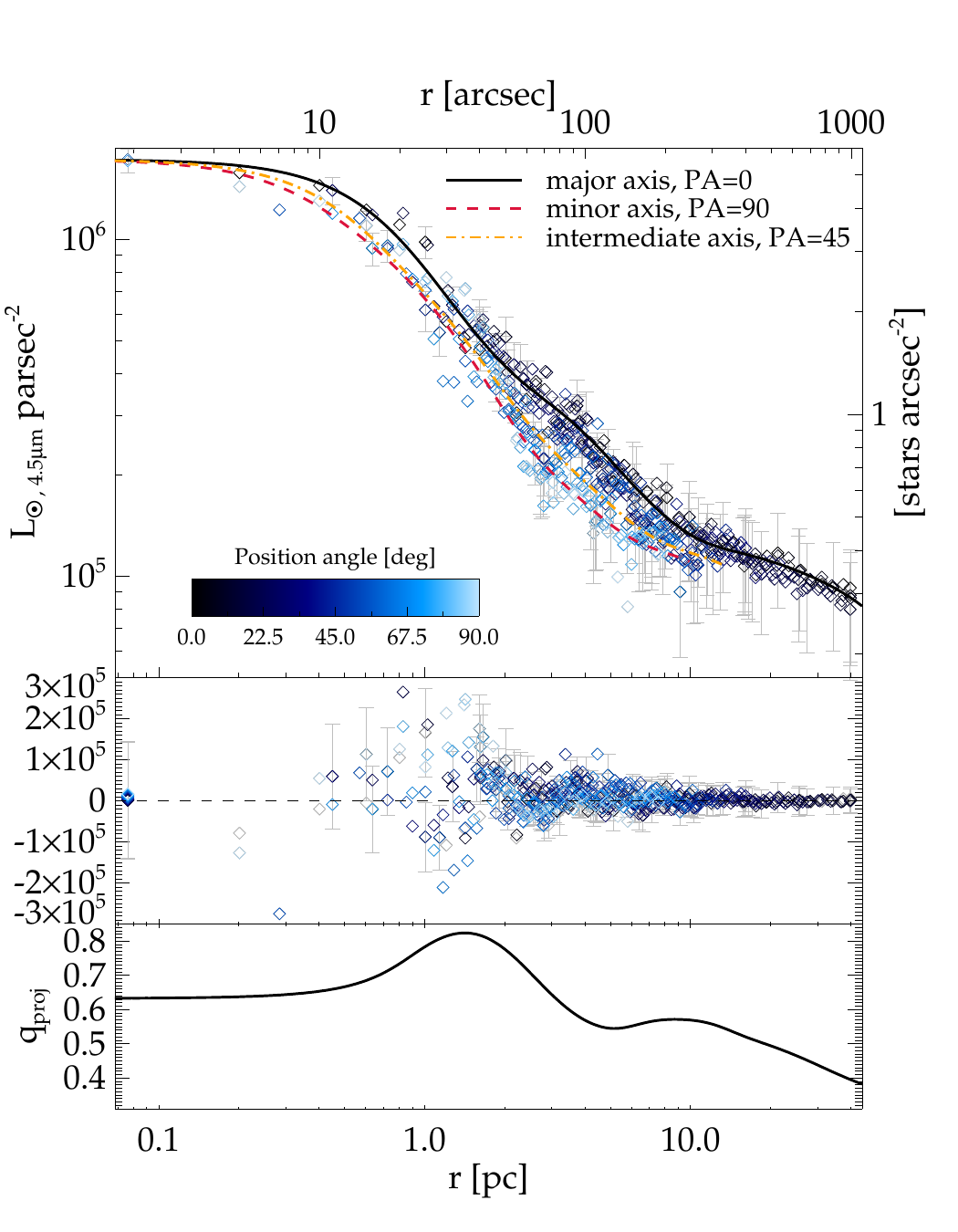} 
	  \caption{Upper panel: Surface density profile derived from the stellar density maps of \citet[blue diamonds]{2020A&A...634A..71G}. The uncertainties are measured using the corresponding uncertainty map and are only shown for a subset of data points to improve visibility.  We matched the profile to the centre value of \cite{2017MNRAS.466.4040F} to convert it into units of \lsun\,pc$^{-2}$. The solid black line denotes the profile along the major axis, the dashed red line shows the profile along the minor axis, and the dot-dashed orange line shows the profile along a position angle of 45\degr. Middle panel: Residuals of the star count data after subtracting the MGE fit.   Lower panel: Projected axis ratio $q_{\rm proj}$. }
  \label{fig:sbdist}
\end{figure}

Finally, we converted the output values of the MGE fit into the values required for the Jeans model, following the equations provided in the \textsc{MgeFit ReadMe}. To convert the central surface brightness into units of \lsun\,pc$^{-2}$, we assumed an exposure time = 1\,s, a zero point =0\,mag, and an extinction $A$=0\,mag, and we adjusted the value of the absolute magnitude for the Sun 
until our density distribution matched the surface brightness distribution of \cite{2017MNRAS.466.4040F} in the 4.5\,\micron\space Spitzer band in the centre.
We list our final stellar surface density MGE in Table \ref{tab:mge} and show the surface density profiles in the top panel of Fig.~\ref{fig:sbdist}. The residuals are shown in the middle panel. They are highest in the central $\lesssim$2\,pc because the density map has some fluctuations caused by the varying extinction. Farther out, the photometry counts are measured in larger regions and combine many pixels,  which produces less scatter and lower residuals. 

\begin{table}
	\centering
	\caption{MGE results from the stellar density map. }
	\label{tab:mge}
	\begin{tabular}{lrcc} 
	\hline
 $l$& $L_l$      & $\sigma_l$ & $q'_l$ \\
   & [$10^4$ \lsun$_{4.5\mu m}$\space pc$^{-2}$] & [arcsec]  & \\ 
 \hline
   1&  44.4012 &  11.315 &  1.000  \\
  2 & 37.6852  & 15.517  & 0.383  \\
  3 & 46.9905  & 22.582  & 1.000  \\
  4 & 16.7216  & 62.626  & 0.450  \\
  5 & 13.3043  & 100.765  & 0.610  \\
  6 & 2.7785  & 285.233  & 0.496  \\
  7 & 11.0686  &1466.691  & 0.300  \\
 \hline
	\end{tabular}
    \tablefoot{The table shows for each Gaussian component the Gaussian number $l$, the central surface brightness $L_l$ (which was matched to the MGE profile of \citealt{2017MNRAS.466.4040F}), the width along the major axis $\sigma_l$, and the projected flattening $q'_l$.}
\end{table}

\section{Discrete dynamical modelling}
\label{sec:model}

We used the discrete Jeans anistotropic MGE (JAM) modelling code by \cite{2013MNRAS.436.2598W}, \textit{CJAM}, which is based on equations and the formalism presented by \cite{2008MNRAS.390...71C}. The basic assumptions are that (i) the distribution function of the stellar system satisfies the collisionless Boltzmann equation, (ii) the model is axisymmetrical, and (iii) the velocity ellipsoid is aligned with cylindrical polar coordinates. 
The model parameters were fitted using \textsc{emcee} \citep{2013PASP..125..306F}, which performed a Markov chain Monte Carlo (MCMC) sampling of the parameter space.

\subsection{Gravitational potential}
\label{sec:pot}
The gravitational potential of the Galactic centre region was modelled by a combination of the supermassive black hole \sgra and the stellar mass distribution. In a subset of models, we also introduced a dark matter (DM) mass distribution. 

For \sgra, we used the well-measured mass of 4.3$\times$10$^6$\,\msun \citep{2022A&A...657L..12G} and modelled it as a single Gaussian with a width of 0\farcs01. This spatial scale is much smaller than the projected distance of the stars in our data set to the position of \sgra, with the shortest being 0\farcs9. Even for the closest star, \sgra therefore produced a Keplerian potential in our model. In one set of models, we did not fix \mbh, but fitted its value.

For the stellar mass distribution, we used the MGE of Table~\ref{tab:mge}, multiplied by a mass conversion factor. This is equivalent to the mass-to-light ratio $\Upsilon$ used in Jeans models that rely on the stellar surface brightness and not on the stellar number density. Because we scaled our MGE to the surface brightness-based MGE of \cite{2017MNRAS.466.4040F}, however, the mass conversion factor is equivalent to a mass-to-light ratio in the 4.5\,\micron band. We expect it to have a value of $\sim$0.6--1.0 \citep{2014ApJ...797...55N,2018MNRAS.473..776K}. 

In most models, we assumed a constant value for $\Upsilon$, but we also tried a radially varying value for $\Upsilon$, parameterised as a modified Osipkov-Merritt profile \citep{1979PAZh....5...77O,1985AJ.....90.1027M}
by 
\begin{equation}
  a(r) = a_0 + \frac{a_\infty - a_0}{1+\left(\frac{R_a}{r}\right)^2},
  \label{eq:2val}
\end{equation}
where $a_0$ is the central value of the quantity (in this case, $\Upsilon$), $a_\infty$ is the asymptotic (outer) value (for $r \rightarrow \infty$), and $R_a$ is the turn-over radius. 

When $\Upsilon$ (or $\beta_z$ or $\kappa$; see Sect. \ref{sec:betakappa}) varied in the model, we needed seven different values for $\Upsilon$, that is, a separate value for each of the seven MGE components. As in \cite{2020MNRAS.492..966K}, we evaluated Eq.~\ref{eq:2val} at the radius $r$ at which a given MGE component contributed most to the combined profile. 
To deproject the surface number density profile, we also required the inclination $i$. We assumed $i$=90\degr, that is, we see the Galactic centre edge-on. 

In addition, we included a DM component in a subset of models. We assumed a spherical dark matter distribution \citep{1996ApJ...462..563N},
\begin{equation}
  \rho_{DM}(r) = \rho_S \left(\frac{r}{r_{DM}}\right)^{-\gamma}\left[1+\left(\frac{r}{r_{DM}}\right)\right]^{-(3-\gamma)}
  \label{eq:nfw},
\end{equation}
where $\rho_S$ is the DM scale density, $r_{DM}$ is the DM scale radius, which we fixed to 8\,kpc (i.e. a value far beyond the extent of our data), and $\gamma$ is the inner DM density slope. When the profile was cored, $\gamma\approx$0, and when the profile was cuspy, $\gamma\approx$1. We ran models for the two cases, $\gamma$=0 and $\gamma$=1, and left only $\rho_S$ as a free parameter. The DM profile was then approximated with a one-dimensional MGE fit. 

\subsection{Anisotropy $\beta_z$ and rotation $\kappa$}
\label{sec:betakappa}
The velocity anisotropy $\beta_z$ and the rotation parameter $\kappa$ describe the velocity structure of the stellar system. 
The anisotropy $\beta_z$ in our models is a measure of the imbalance of radial motion and vertical motion. The models assume a cylindrically aligned velocity ellipsoid (i.e. $\overline{v_R v_z} =0$\footnote{$\overline{v_R v_z}$ denotes the cross-term of the second velocity tensor, along the cylindrical radial $R$ direction and the z-direction.}), and $\beta_z$ is defined as $\beta_z=1-{\overline{v_z^2}}/{\overline{v_R^2}}$ in the meridional plane\footnote{$\overline{v_z^2}$ denotes the second velocity moment in the z-direction, and it indictates $\overline{v_R^2}$ in the cylindrical radial $R$ direction.}.
When $\beta_z$=0, there is no imbalance, and the velocity structure of the model is isotropic. For $\beta_z$\textless0, there is a tangential anisotropy, and for $\beta_z$\textgreater0, there is a radial anisotropy, and the stellar orbits are more radial.

It is possible to assume a radially constant value for $\beta_z$ or to let it vary for each stellar MGE component. In some models, we parameterised $\beta_z(r)$ via Eq.~\ref{eq:2val} (i.e. three parameters: $\beta_0, \beta_\infty$, and $R_\beta$).

The rotation parameter $\kappa$ quantifies the amount of rotation of the system. $\kappa$ is defined as $\kappa=\overline{v_\phi}/(\surd{(\overline{v_\phi^2}-\overline{v_R^2})}$, and it is a measure of the ratio of ordered to disordered motion\footnote{$\overline{v_\phi}$ and $\overline{v_\phi^2}$ denote the first and second velocity moment along the tangential cylindrical coordinate $\phi$.}. When $\kappa$=0, there is no rotation. An isotropic rotator would be described by $\beta_z$=0 and $\kappa=\pm$1. The sign of $\kappa$ denotes the direction of the rotation.  In the Galactic centre, we expect significant rotation and  a negative value for $\kappa$ (based on the definition of the coordinate system by \citealt{2013MNRAS.436.2598W}). A positive value would mean that the sign of the angular momentum vector was reversed.

As with $\beta$, it is possible to implement different $\kappa$ parametrisations: A constant value of $\kappa$ (i.e. the same $\kappa$ for each stellar MGE component, one additional parameter), or $\kappa$ as a function of radius, parametrised via Eq.~\ref{eq:2val} (i.e. three parameters: $\kappa_0, \kappa_\infty$, and $R_\kappa$). This three-parameter model allowed us to change the rotation from the NSC to the NSD.

\subsection{Population components}
It is possible to build multi-population models with discrete posterior distribution functions, as shown for example by \cite{2016MNRAS.462.4001Z,2016MNRAS.463.1117Z,2020MNRAS.492..966K,2022ApJ...939..118K}, and we followed a similar procedure. We either included one or two population components and also a background component that represented the Galactic bar. In principle, it is possible to include even more components, but this would require more precise \mh data or additional data, such as elemental abundances or reliable colours, which are not available for a sufficiently large sample in the Galactic centre. 

We solved the Jeans equations for both components in the same gravitational potential, but with a different tracer density $I$, $\beta_z$, and $\kappa$ for the two components. We estimated the probability of each star in our kinematic sample to belong to one of the modelled stellar populations $(k)$ or a background component ($bg$) given its on-sky position, velocity (\vlos and, if available, PMs), and chemical population tracer (\mh). 
In summary, there were up to three probabilities for each star: the dynamical probability, the spatial probability, and (when $k$>1) the chemical probability.

\subsection{Dynamical probability}
Given the gravitational potential, $\beta_z$, and $\kappa$, \textit{CJAM} solves the Jeans equations and predicts the first and second velocity moments in three directions $j$ and at the location of each star $i$ in our sample. The first moment is the mean velocity $\mu$, and the second moment is the quadratic sum of the mean velocity $\mu$ and the velocity dispersion $\sigma$.

Based on the measured velocity $V_{\text{LOS},i} \pm \delta V_{\text{LOS},i}$, and for some stars, also the proper motions $V_{l,i} \pm \delta V_{l,i}$ and $V_{b,i} \pm \delta V_{b,i}$, we computed the dynamical probability of star $i$ to belong to population $k$ in one dimension $j$,

\begin{equation}
P_{\text{dyn},j,i}^k=\frac{1}{\sqrt{(\sigma_{j,i}^k)^2 + (\delta V_{j,i})^2}}\cdot\exp{\left[-\frac{1}{2}\frac{(V_{j,i}-\mu_{j,i}^k)^2}{(\sigma_{j,i}^k)^2+(\delta V_{j,i})^2}\right]}, 
\label{eq:pdyn}
\end{equation}
where $\mu^k_{j,i}$ and $\sigma_{j,i}^k$ are the velocity and velocity dispersion predicted by the dynamical model of population $k$ at the sky position of star $i$, either along the line of sight, parallel to the Galactic plane ($l$), or perpendicular to it ($b$). When a star only had \vlos, $P_{\text{dyn},i}^k=P_{\text{dyn},j,i}^k$. When star $i$ also had
proper motions, the dynamical probability was $P_{\text{dyn},i}^k=\prod_{j=1}^3P_{\text{dyn},j,i}^k$. 

For the background component, we did not solve the Jeans equations, but assumed a Gaussian velocity distribution with a fixed 
 mean velocity $\mu^{\rm bg}$=0\,\kms at the location of each star $i$ and in each direction $j$ and a velocity dispersion of $\sigma^{\rm bg}$=130\,\kms\ \citep{2017MNRAS.465.1621P} that is consistent with the Galactic bar.

\subsection{Spatial probability}
We assumed that our MGE (Table ~\ref{tab:mge}) represented the total stellar surface density distribution $I$ in the Galactic centre. For two populations, their densities $I^k$ add up to $I$. The fraction $h^k$ contributed by population $k$ was able to vary as a function of radius, and we parametrised the change in $h^k$ with radius with Eq.~\ref{eq:2val}, that is, with a monotonous three-parameter function. 
Because we had only two populations, we had $I^1(x_i,y_i)= h (r) \cdot I(x_i,y_i)$ and $I^2(x_i,y_i)= (1 - h (r)) \cdot I(x_i,y_i)$. 
We used the respective $I^k$ to solve the Jeans equations for the two populations separately. 

We assumed that the background surface number density is uniform, which is a fair assumption given the small extent of the modelled region in comparison to the size of the bar. We introduced the parameter $\epsilon$, which denotes the fraction of contaminant stars with respect to the central surface density $I(0,0)$.

The spatial probability for star $i$ at the position $(x_i,y_i)$ to belong to population $k$ is 
\begin{equation}
P_{\text{spa},i}^k=\frac{I^k(x_i,y_i)}{I(x_i,y_i)+I_{bg}}=\frac{I^k(x_i,y_i)}{I(x_i,y_i)+\epsilon\cdot I(0,0)}, 
\label{eq:pspa}
\end{equation}
and the spatial probability for star $i$ to belong to the contaminant population is simply
\begin{equation}
P_{\text{spa},i}^{bg}=1- \sum_k P_{\text{spa},i}^k. 
\label{eq:pspa2}
\end{equation}

\subsection{Chemical probability}
The NSC and NSD may contain stars that belong to different stellar populations with different dynamical properties ($\beta_z, \kappa$). This means that stars may have formed at different times and from different materials. This may be reflected in the stellar metallicity \mh, and we can use it as a stellar population tracer.

We assumed that each population $k$ has a Gaussian metallicity distribution with a mean metallicity $Z^k_0$ and a metallicity dispersion $\sigma^k_Z$. For star $i$ with a measured metallicity $Z_i \pm \delta Z_i$, the chemical probability for population $k$ becomes 
\begin{equation}
P_{\text{chm},i}^k = \frac{1}{\sqrt{2\pi\left[(\sigma^k_Z)^2+(\delta Z_i)^2)\right]}}\cdot \exp{\left[-\frac{1}{2}\frac{(Z_i-Z^k_0)^2}{(\sigma^k_Z)^2+(\delta Z_i)^2} \right]}.
\label{eq:pchem}
\end{equation}

We used the overall metallicity \mh and its uncertainty derived from full spectral fitting. For the background component, we used a Gaussian distribution centred on $Z^{bg}_0$=0.07\,dex and with a width of $\sigma_{Z}^{bg}$=0.3\,dex, as found by \cite{2021A&A...649A..83F} on a bulge control field observed with KMOS.

\subsection{Total probability} 

The total probability for star $i$ in a given model is then 

\begin{equation}
L_{i} = \left(\sum_{k\neq bg} P_{\text{spa},i}^k\cdot P_{\text{chm},i}^k\cdot P_{\text{dyn},i}^k\right) + P_{\text{spa},i}^{bg}\cdot P_{\text{chm},i}^{bg}\cdot P_{\text{dyn},i}^{bg}, 
\label{eq:toti}
\end{equation}
where $k$ denotes the two populations that are not the background. The total likelihood $L$ of a model is given by 
\begin{equation}
L = \prod_{i=1}^N L_i,
\label{eq:tot}
\end{equation}
where $i$ runs over all $N$ stars.

For a given star, its likelihood of belonging to population $k$ is $L_{i}^k = P_{\text{spa},i}^k\cdot P_{\text{chm},i}^k\cdot P_{\text{dyn},i}^k$, and the probability of belonging to a given population is $P_{i}^k=L_{i}^k/(L_i^{bg}+\sum_k L_{i}^k)$. 
When we fit only one population, we only have the probability for a single population and the background, and we still lack the chemical probability ($P_{\text{chm},i}^k=P_{\text{chm},i}^{bg}$=1).

\subsection{One-population parameter summary}

We ran several sets of models with only one population and the background component, where we only considered the dynamical and spatial probability. These models had the parameters listed below.
\begin{enumerate}[label=(\roman*),nosep]
\item The background percentage parameter $\epsilon$. This is one parameter, given in percent, and has a uniform prior  [0, 10].
\item The anisotropy $\beta_z$. This is three parameters, and the uniform priors $\beta_{0,\infty}$ are [-1.5, 0.9], $\log(R_\beta)$=[1.079, 2.91], and $R_\beta$ in arcseconds.
\item The rotation $\kappa$. This is three parameters with uniform priors $\kappa_{0,\infty}$ of [-1.5, 0.4], $\log(R_\kappa)$=[1.079, 2.91], and $R_\kappa$ in arcseconds.
\end{enumerate}
Each run focused on a different part of the gravitational potential, as listed below.
\begin{enumerate}[label=(\roman*),nosep]
\item On the mass of the supermassive black hole \mbh. This is one parameter, and the uniform prior is [2, 6]$\times$10$^6$\,\msun.
\item On the mass-to-light conversion $\Upsilon$. This is three parameters, and the uniform priors $\Upsilon_{0, \infty}$ are [0.1, 3.0], $\log(R_\Upsilon)$=[1.079, 2.91], and $R_\Upsilon$ in arcseconds.
\item On the dark matter scale density $\rho_S$. This is one parameter, and it is in the range $\log(\rho_S)$=[-5, 5].

\end{enumerate}

\noindent We always included a single value for $\epsilon$ and a modified Osipkov-Merrit profile (Eq.~\ref{eq:2val}) for $\beta_z$ and $\kappa$. We also used $\Upsilon$ in each model, but implemented different parametrisations. In Sections \ref{sec:fitbh} and \ref{sec:fitdm}, we use a constant value for $\Upsilon$, that is, we used a single parameter that did not vary with the radius, and we either included \mbh (nine free parameters in total) or $\rho_S$ (nine free parameters in total) in the fit. In Sect. \ref{sec:fitmlr} we fix \mbh, neglect DM, and fit a radially varying $\Upsilon$ parametrised with a modified Osipkov-Merrit profile (ten free parameters in total; see also Table ~\ref{tab:pop1bg}). 
The bounds of the scale radii ($R_\beta, R_\kappa, R_\Upsilon$) were chosen to be larger than the innermost MGE width $\sigma$ and not larger than the outermost discrete kinematic measurement.

We started the chains at $\beta_{0,\infty}$=0 (i.e. isotropy), with moderate rotation ($\kappa_0$=-0.9, $\kappa_\infty$=-0.5) with a transition at $R_{\beta,\kappa,\Upsilon}$= 45\,arcsec, $\Upsilon$=1, $\epsilon$=5\%, \mbh=4.3$\times$10$^6$\,\msun, and $\log(\rho_S)$=0.
For each model set-up, we ran \textsc{emcee} \citep{2013PASP..125..306F} with 100 walkers and at least 1\,900 steps, and we discarded the first 20\% of the steps before the chains converged. We also tested longer chains for the non-DM models with more than 6\,000 steps, but our best-fit parameters did not change significantly.

\subsection{Two-population parameter summary}

In the two-population model, we considered the total probability and a background component. The parameters are listed below.
\begin{enumerate}[label=(\roman*),nosep]
\item The background percentage parameter $\epsilon$. This is one parameter, given in percent, and has a uniform prior  [0, 10].
\item The anisotropy $\beta_z^k$. These are two parameters, each with a uniform prior range [-1.5, 0.9] and constant with radius, for $k$=2 populations.
\item The rotation $\kappa^k$. These are two parameters, each with a  uniform prior range [-1.5, 0.4] and constant with radius,  for $k$=2 populations.
\item The mass-to-light conversion $\Upsilon$. This is one parameter with a uniform prior range [0.1, 3.0] and constant with radius. We use the same $\Upsilon$  for both populations $k$.
\item The population fraction $h$. This is  three parameters with uniform priors $h_{0,\infty}$ of [0.5, 1.0], $\log(R_h)$=[1.079,2.91],  and $R_h$ in arcseconds. 
\item The Gaussian mean metallicity $Z_0^k$ and dispersion $\sigma_Z^k$. These are  four parameters, and the uniform priors are $Z_0^1$=[0.0, 0.8]\,dex, $\sigma_Z^1$=[0.2, 1.5]\,dex, $Z_0^2$ =[-1.6, 0.2]\,dex, and $\sigma_Z^2$=[0.2, 1.5]\,dex.

\end{enumerate}
\noindent  We fit a constant value for $\epsilon$, while the population fraction $h$ was parametrised with Eq.~\ref{eq:2val}.
We started the chains at $\beta_z^k$=0 (i.e. isotropy), moderate rotation ($\kappa^{k}$=-0.8), $Z_0^1$=0.35\,dex, $Z_0^2$=-0.6\,dex, $\sigma_Z^{k}$=0.4\,dex, $h_{0,\infty}=0.75$, $R_h$=100\,arcsec, $\epsilon$=2.5\%, and $\Upsilon$=0.71.
We had 13 parameters in total. For this model set-up, we ran \textsc{emcee} with 100 walkers, 4\,700 steps, and a burn-in fraction of 0.3.

\label{sec:fit}
\section{One-population dynamical models}
\label{sec:result}

\begin{table}
\centering
\caption{Results of one-population dynamical models with a radial varying $\beta_z$ and $\kappa$ and with various gravitational components that were fitted. }
 \label{tab:pop1bg}
%\begin{tabular}{lcccc}
\begin{tabular}{l@{\hskip 4pt}c@{\hskip 4pt}c@{\hskip 4pt}c@{\hskip 4pt}c}
 \hline
 \noalign{\smallskip}

Parameter& Free \mbh   &  $\Upsilon$(r)&DM cusp &DM core    \\   
&       &&$\gamma$=1   & $\gamma$=0 \\  
 &Sect. \ref{sec:fitbh}& Sect. \ref{sec:fitmlr}& Sect. \ref{sec:fitdm}& Sect. \ref{sec:fitdm}  \\ 
\hline
\noalign{\smallskip}
$\epsilon$    &2.4 $\pm$0.2  &2.4 $\pm$0.2        &2.4 $\pm$0.2         &2.4 $\pm$0.2        \\ 
\noalign{\smallskip}
$\beta_0$       &-0.15 $^{+0.11}_{-0.19}$ &-0.13 $^{+0.11}_{-0.18}$  &-0.23 $^{+0.15}_{-0.31}$  &-0.22 $^{+0.14}_{-0.27}$   \\ 
\noalign{\smallskip}
$\beta_\infty$     &-0.15 $^{+0.08}_{-0.09}$   &-0.15 $^{+0.09}_{-0.10}$ & -0.07 $\pm$0.10   &-0.04 $\pm$0.10  \\ 
\noalign{\smallskip}
$R_\beta$  &58 $^{+275}_{-41}$   &68 $^{+309}_{-50}$   &53 $^{+322}_{-36}$     &58 $^{+324}_{-41}$        \\ 
\noalign{\smallskip}
$\kappa_0$       &-0.26 $^{+0.35}_{-0.33}$ &-0.32 $^{+0.39}_{-0.38}$     &-0.26 $^{+0.38}_{-0.34}$  &-0.23 $^{+0.38}_{-0.35}$  \\ 
\noalign{\smallskip}
$\kappa_\infty$     &-1.10 $\pm$0.05  &-1.09 $^{+0.05}_{-0.06}$   &-1.08 $\pm$0.05      &-1.11 $^{+0.05}_{-0.06}$     \\ 
\noalign{\smallskip}
$R_\kappa$  &17 $^{+8}_{-4}$     &17 $^{+10}_{-4}$    &17 $^{+8}_{-4}$      &17 $^{+9}_{-4}$        \\ 
\noalign{\smallskip}
\hline
\noalign{\smallskip}
$\Upsilon_0$     & ...      &0.90 $^{+0.28}_{-0.12}$       &...             &...             \\ 
\noalign{\smallskip}
$\Upsilon_\infty$     &...       &0.68 $\pm$0.04          &...             &...           \\ 
\noalign{\smallskip}
$R_\Upsilon$ &...     &51 $^{+280}_{-35}$        &...             &...              \\ 
\noalign{\smallskip}
$\Upsilon$       &0.72$\pm$0.03    &...    &0.60 $^{+0.08}_{-0.06}$   &0.62 $^{+0.07}_{-0.05}$     \\ 
\noalign{\smallskip}
\mbh  &4.35$^{+0.24}_{-0.23}$    & (4.3)           & (4.3)       & (4.3)                \\
\noalign{\smallskip}
$\log(\rho_S)$   &...        &...       &-0.57 $^{+0.16}_{-0.37}$    &1.62$^{+0.15}_{-0.56}$     \\ 
\noalign{\smallskip}
\hline 
\noalign{\smallskip}
Number \\ 
of free & 9 & 10 & 9 &9\\
parameters \\
\noalign{\smallskip}
\hline 
\end{tabular}
\tablefoot{The listed values are the medians of the posterior distributions, and the uncertainties are the 16th and 84th percentiles. The columns refer to the results in Sects. \ref{sec:fitbh}, \ref{sec:fitmlr}, and \ref{sec:fitdm}. The units of $R_\beta, R_\kappa, and R_\Upsilon$ are in arcsec, \mbh in is given 10$^6$\,$M_\odot$, and $\epsilon$ is listed in percent}
\end{table}

In this section, we describe the results we obtained when we only fitted the stellar kinematic data and did not try to separate the data into two populations.  We ran several sets of models that tested   whether our results were consistent with the known \mbh mass (Sect. \ref{sec:fitbh}), if $\Upsilon$ varied with radius (Sect. \ref{sec:fitmlr}), and if a DM component was necessary (Sect. \ref{sec:fitdm}). 

The results of the runs are summarised in Table \ref{tab:pop1bg}. The percentage of background stars $\epsilon$, the velocity anisotropy $\beta_z$, and the rotation parameter $\kappa$ did not alter significantly when we changed the components that contribute to the gravitational potential. The results are consistent because the DM contribution is low, and our \mbh result agrees with the reference value.  We use this knowledge for the two-population dynamical models in Sect. \ref{sec:res2pop}.

\subsection{Fitting $M_\bullet$}
\label{sec:fitbh}
The mass of \sgra, \mbh, is well known to be (4.30$\pm$ 0.012)$\times$10$^6$\,\msun \citep{2022A&A...657L..12G}. We tested whether our data were consistent with this value by fitting the value of \mbh.

In this fit with a radially constant $\Upsilon$, our value for \mbh = $(4.35^{+0.24}_{-0.23}) \times$10$^6$\,\msun is remarkably close to the stellar orbit measurements reported for instance by \cite{2016ApJ...830...17B,2022A&A...657L..12G}. 
The mass of the supermassive black hole, \sgra, has been attempted to be determined numerous times via dynamical models. The results from the stellar orbits often do not match, and the mass of \sgra is often underestimated with dynamical models \citep[see e.g.][and references therein]{2019MNRAS.484.1166M}. 
Our result is close to the reference value. This shows us that our data do not disagree with \mbh=4.3$\times$10$^6$\,\msun, and hence, they cause no biases in the other parameter fits. We therefore fixed \mbh to the literature value in all future runs. We show the corner plot of the posterior distribution in Fig.~\ref{fig:cornerbh}. 

\subsection{Radially varying $\Upsilon$}
\label{sec:fitmlr}

We tested whether the conversion of the stellar number density (which traces the stellar light) into the stellar mass is constant as a function of radius.  We parametrised $\Upsilon$ with an Osipkov-Merritt profile (Eq. \ref{eq:2val}), which means that there is an inner value $\Upsilon_0$, an outer value $\Upsilon_\infty$, and a transition radius $R_\Upsilon$. 

\begin{figure}
\centering
 \includegraphics[width=0.95\columnwidth]{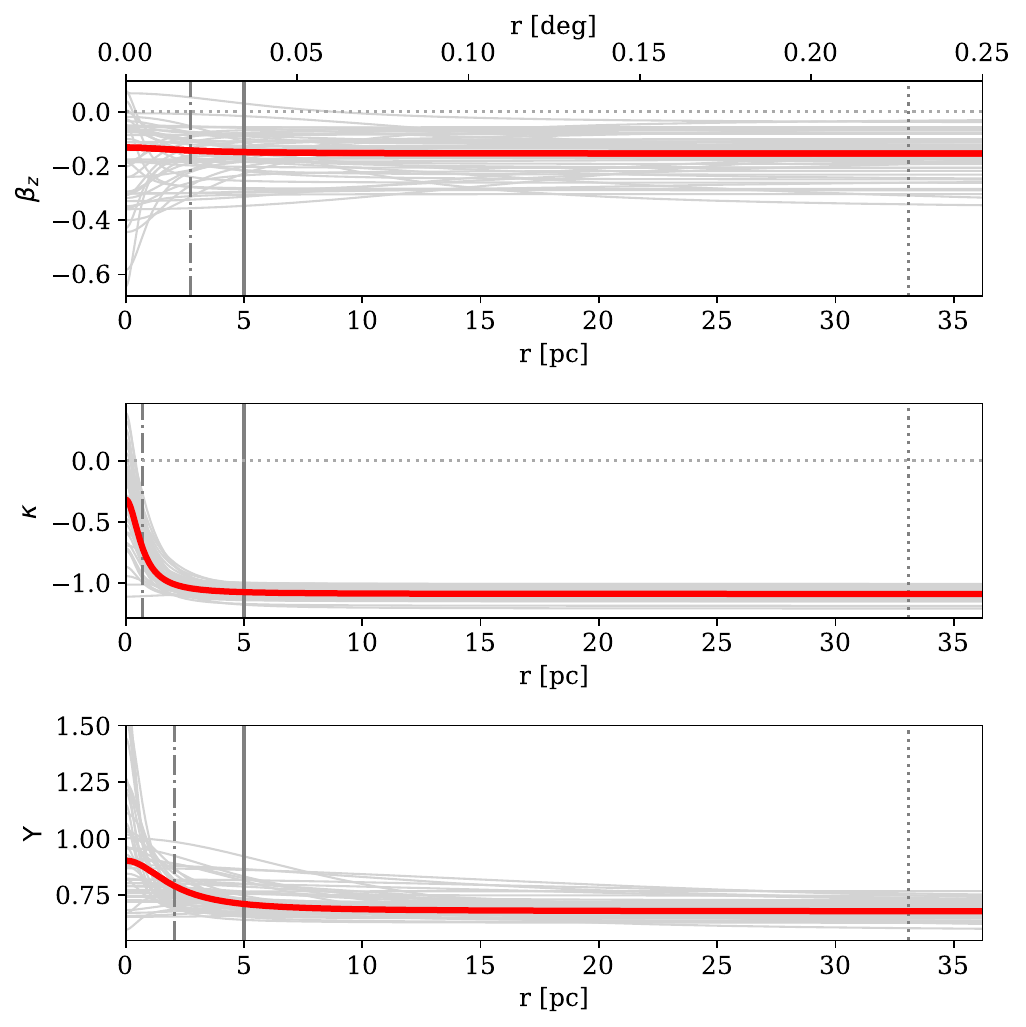}
  \caption{Radial profile of the velocity anisotropy $\beta_z$ (top), the rotation $\kappa$ (middle), and 
  the mass-to-number density conversion $\Upsilon$ (bottom) as derived in Sect. \ref{sec:fitmlr} using a three-parameter function. The red lines denote the median of the posterior distribution, and the grey lines are 50 randomly drawn realisations from the posterior distribution. The vertical solid lines denote 1\,\re=5\,pc, the dotted lines show the outer limit of the kinematic data, and the dot-dashed lines show the median of $R_\beta$, $R_\kappa$, and $R_\Upsilon$, respectively.
}
  \label{fig:betakappaml}
\end{figure}

The inner value of $\Upsilon_0$ is higher than the outer value $\Upsilon_\infty$, but because of the uncertainties, the difference is small (smaller than 1.5 times the sum of the uncertainties). The value of $\Upsilon_\infty$ is very well constrained, with much smaller uncertainties than for $\Upsilon_0$. The radius at which the transition occurs also has a large upper uncertainty, and its 1$\sigma$ upper limit is at $\sim$13\,pc. We show the $\Upsilon$ profile in the bottom panel of Fig.~\ref{fig:betakappaml}. 
The radially constant $\Upsilon$ that we fitted in the previous section lies in between our $\Upsilon_0$, and $\Upsilon_\infty$, and it is consistent with $\Upsilon_\infty$ within the uncertainties. This suggests that the firm constraints on $\Upsilon$ come from data in the outer region.
The corner plot of the posterior distribution is shown in Fig.~\ref{fig:cornerml3}.

The mass-to-light ratio in the mid-infrared is rather constant compared to other bands, and variations with age or metallicity are small, 
with only a factor of two for ages between 3 and 10\,Gyr \citep{2014ApJ...788..144M}. The values of 0.6-0.9 are consistent with ages of 5--10\,Gyr and with a sub-solar to super-solar metallicity \citep{2014ApJ...797...55N}. Older stellar populations and populations with lower metallicity tend to have a higher $\Upsilon$. 
This means that the higher $\Upsilon_0$ agrees with either an older population or with a lower [Fe/H] compared to $\Upsilon_\infty$. A high concentration of dark remnants (stellar mass black holes or neutron stars) around \sgra may also increase $\Upsilon_0$ compared to $\Upsilon_\infty$, however.

In summary, the profile $\Upsilon$ only shows a slight radial dependence that only affects the innermost $r\lesssim$2\,pc, and all other parameters are robust. We therefore assumed that $\Upsilon$ is constant in our DM and two-population models. 

\subsection{Dark matter scale density} 
\label{sec:fitdm}
We tested whether we needed to include DM in our models by fitting Jeans models with either a DM cusp ($\gamma=1$) or a DM core ($\gamma=0$). We only fitted the DM scale density $\rho_S$. We assumed that $\Upsilon$ is constant with radius, and we fixed \mbh to the literature value.

For DM cusp and DM core, the upper value of $\rho_S$ is more tightly constrained than the lower value, especially for the cored DM profile. For both DM profiles, the enclosed mass assigned to DM within 1\,\re of the NSC is significantly below the stellar mass. In particular, the enclosed DM mass for a DM core profile at deprojected radius $r$=5\,pc=1\,\re is below 0.5\% (84th percentile upper limit) and lower than $\sim$12\% of the total mass at 33\,pc (the outer extent of the kinematic data). For a DM cusp model, the enclosed mass of DM is higher, at $\sim$1--4 \% of the total enclosed mass at 5\,pc (16th and 84th percentile limits) and $\sim$6--25\% at 33\,pc.  The DM scale radius in these models was fixed to be much farther out than our data ($r_{DM}$=8\,kpc). We tried a higher value ($r_{DM}$=17.2\,kpc), and the DM mass fractions were in the same ranges. We show the corner plots of the posterior distributions in Figs.~\ref{fig:cornerdmcusp}-\ref{fig:cornerdmcore}.

Simulations of Milky Way analogues appear to favour an inner logarithmic density slope of 0.5 to 1.3 and exclude a cored DM profile  \citep{2025arXiv250114868H}.
Our assumed DM profiles of $\gamma$=1 may still not reflect the true distribution.  Close to \sgra, gravitational encounters, self-annihilation, or scattering may alter the density profile within $r$\textless 1\,pc \citep{Merritt_2010}. These processes are not likely to affect the outer regions, however. 
We tested two extreme cases for the DM profile, and both resulted in a small contribution to the total mass in the region of our data. We note that several model parameters ($\epsilon, \beta_0, \beta_\infty, R_\beta, \kappa_0, \kappa_\infty, R_\kappa$) changed only slightly compared to the models without DM and within their uncertainties. The only exception was $\Upsilon$, which changed by more than $\sim$1$\sigma$ for the cusp profile. Because most parameters were barely affected by the inclusion of a DM profile, we did not include a DM component for the two-population models because the values of $\rho_S$ and $\gamma$ are poorly constrained and the computing time increased significantly when we included DM.

\subsection{Model properties}
\begin{figure}
\centering
 \includegraphics[width=0.95\columnwidth]{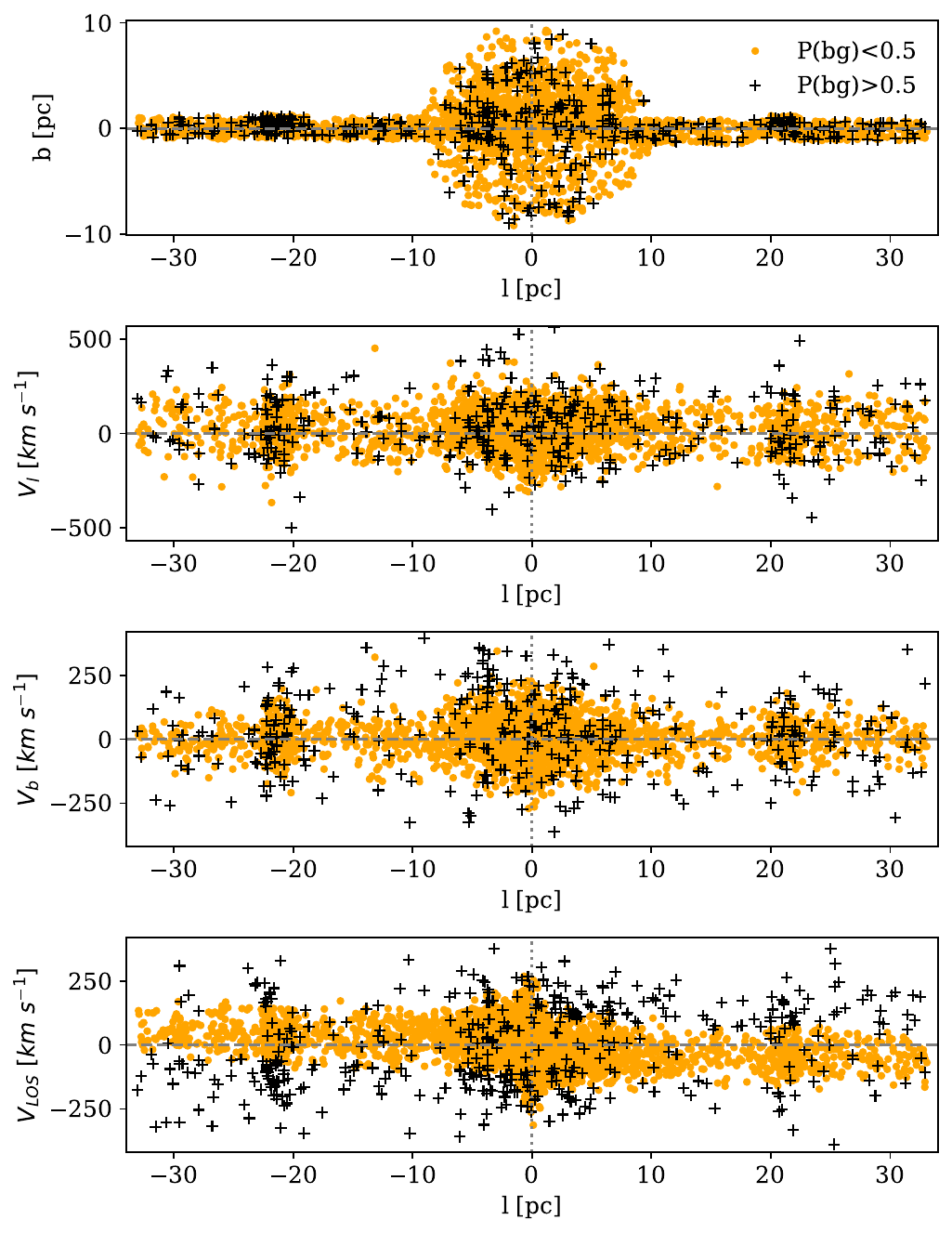}
  \caption{Map of the stellar positions (top) and position-velocity plots along the Galactic longitude $l$ for the PM along $l$ (second panel), along $b$ (third panel), and along the line of sight (fourth panel). The colour-coding is from the median realisation of the one-population model in Sect. \ref{sec:fitmlr}. Stars with a probability to be a member star higher than 0.5 are shown as an orange circle, and background stars are shown as a black pluses. 
  }
  \label{fig:bgcontr}
\end{figure}

Our one-population models give fully consistent results for the fraction of background stars $\epsilon$, the velocity anisotropy $\beta_z$, and the rotation parameter $\kappa$. $\epsilon$ is consistently at 2.4\%, which corresponds to $\sim$450 stars with a probability $P_i^{bg}>0.5$. We show their spatial distribution and location in a position-velocity diagram in Fig.~\ref{fig:bgcontr} using the 50th percentile realisation of Sect.~\ref{sec:fitmlr} with a radially varying $\Upsilon$ as an example. Other models gave similar results. The background stars were identified via high velocities in the PM or \vlos component. Because only $\sim$75\% of the stars in our catalogue have PM measurements, the value of $\epsilon$ may be underestimated.

The models found a mild tangential anisotropy, with $\beta_z\sim$--0.1 to --0.2. They did not find a strong variation in the anisotropy with radius. $\beta_0$ and $\beta_\infty$ differed by less than their uncertainties, and $R_\beta$ was poorly constrained. We conclude that using a constant value for $\beta_z$ is a fair assumption. An even more complicated $\beta_z$ profile such as the profile used by \cite{2013ApJ...779L...6D}, with a fourth parameter to adjust the sharpness of the transition from $\beta_0$ to $\beta_\infty$, is not required to describe our kinematic data. With a larger data set or more precise kinematic data, a more complicated functional form for $\beta_z$ would  be worth investigating. We show the $\beta_z$ profiles drawn from the posterior distribution of Sect.~\ref{sec:fitmlr} in the top panel of Fig.~\ref{fig:betakappaml} as an example. The profiles were very similar in the other runs. 

The rotation parameter $\kappa$ is also consistent among the different models, and we show an example from Sect.~\ref{sec:fitmlr} in the middle panel of Fig.~\ref{fig:betakappaml}. The radial change from $\kappa_0$ to $\kappa_\infty$ indicates a lower ratio of the ordered to disordered motion in the centre than farther out. This change occurs within the central $\lesssim$1.2\,pc, where  \sgra dominates the gravitational potential. The value of $\kappa_0$ is even consistent with zero in the centre within the 84th percentile. This only holds for the inner $\sim$1\,pc. There is strong and significant rotational support at larger radii. 
The innermost 1\,pc may have different dynamical properties, which may be caused by the influence of \sgra. 
The velocity anisotropy or rotation parameters in the region of the inner few parsecs and the outer regions do not differ much, however, and we assumed that they are constant with radius in the two-population models. 

\begin{figure}
\centering
 \includegraphics[width=0.95\columnwidth]{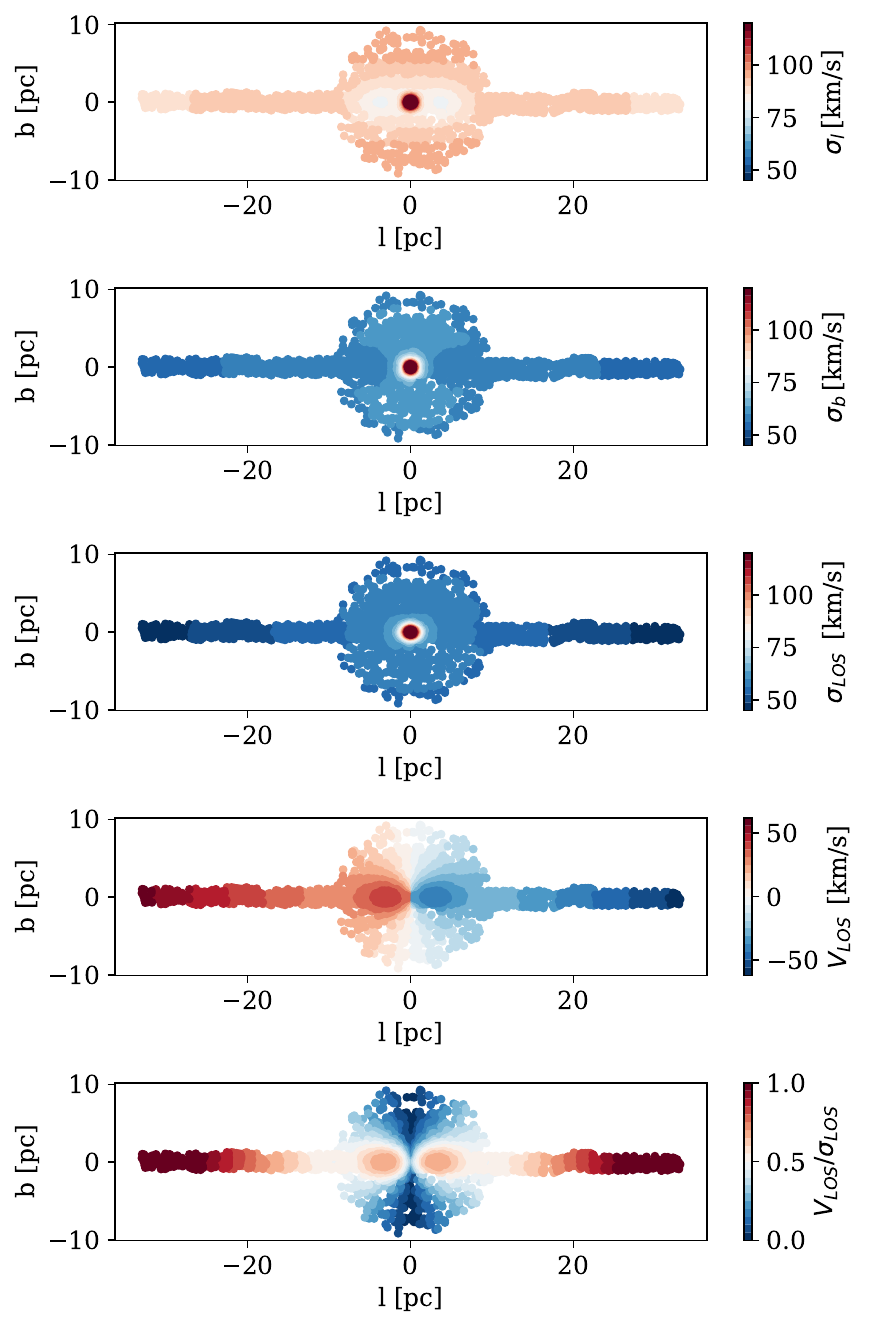}
  \caption{Map of the velocity dispersion in three directions ($\sigma_l$, $\sigma_b$, and \slos), \vlos, and \vlos/\slos for the median model with a radially varying $\Upsilon$. 
  }
  \label{fig:vlos1p}
\end{figure}

The velocity dispersion maps in three dimensions (along the Galactic longitude $l$, along the latitude $b$, and along the line of sight), the \vlos map, and \vlos/\slos maps of the same model in Fig.~\ref{fig:vlos1p} show us that within the region of the NSC, there are distinct peaks in \vlos ($\sim\pm$41\,\kms) and \vlos/\slos that are followed by a drop at $l\sim$10\,pc. $\vert$\vlos$\vert$ and \vlos/\slos increase again at larger $l$. The peak value within the NSC is at \vlos/\slos=0.68, the minimum is at 0.47, and it then continues to rise further to \vlos/\slos\textgreater 1.2 and \vlos$\sim\pm$58\,\kms at $l\sim$30\,pc.

\begin{figure*}
  \includegraphics[width=\columnwidth]{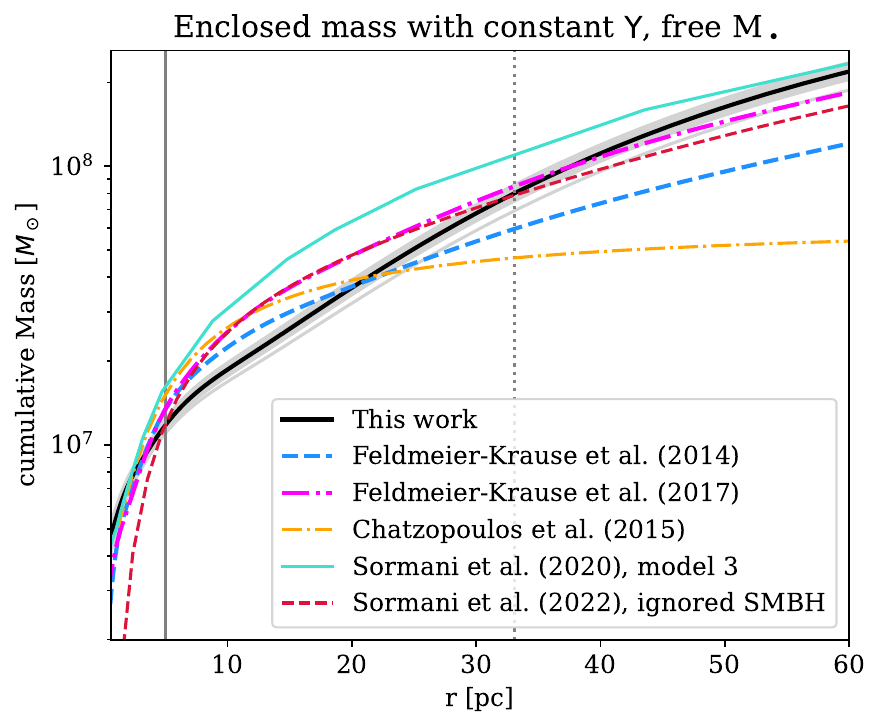}
  \includegraphics[width=\columnwidth]{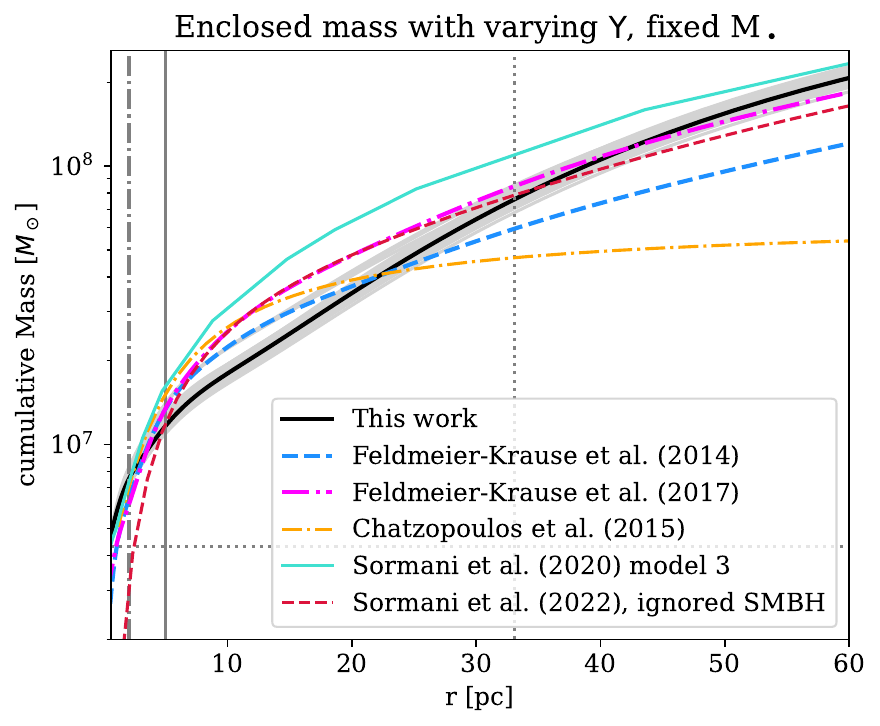}
  \includegraphics[width=\columnwidth]{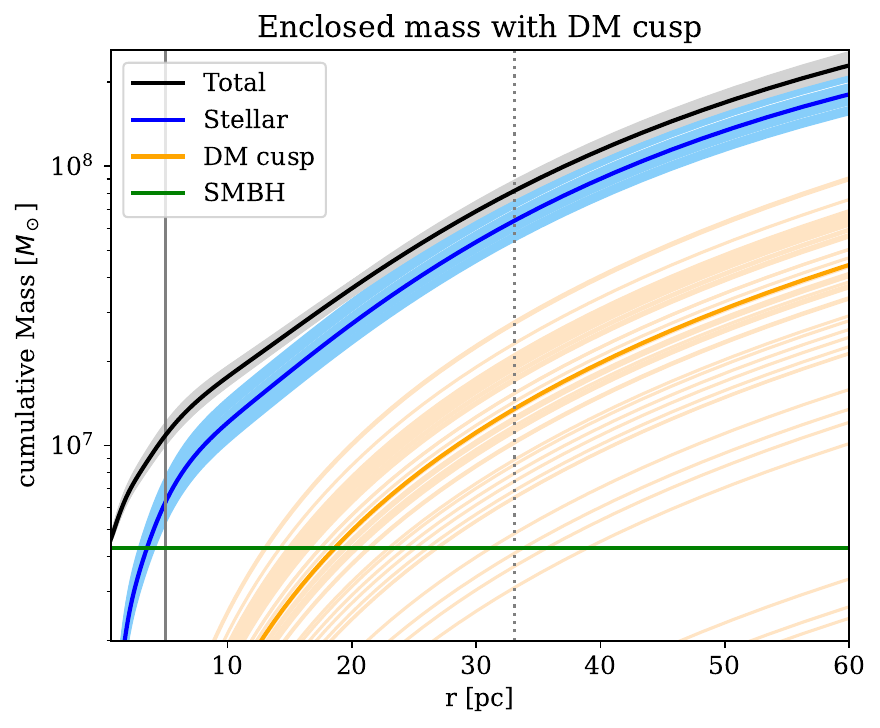}
  \includegraphics[width=\columnwidth]{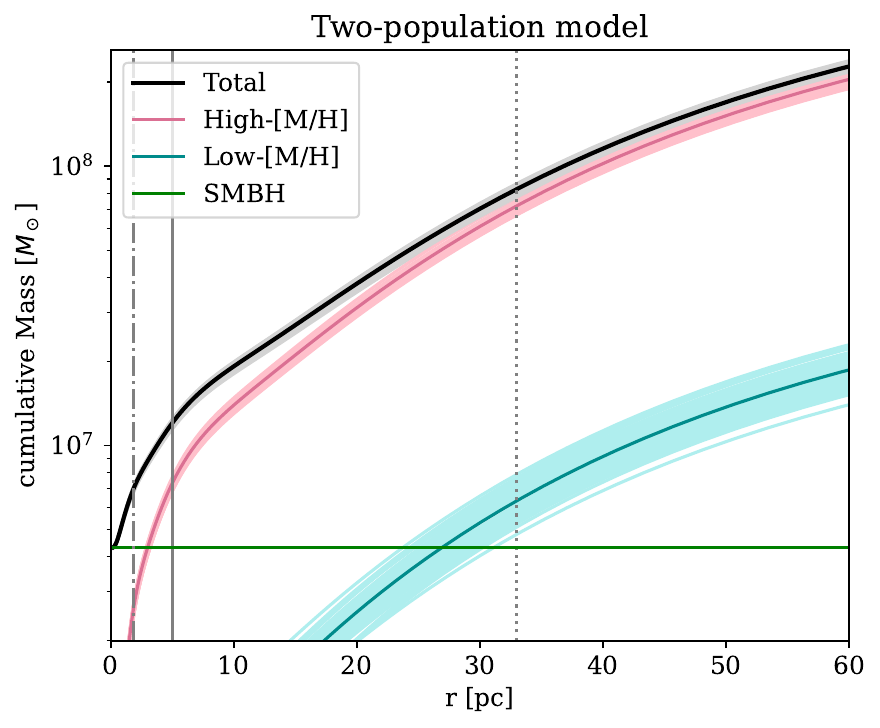}
  \caption{Total enclosed mass as a function of spherical deprojected radius for the one-population fits with free \mbh\ (top left), radially varying $\Upsilon$ (top right), a DM cusp (bottom left), and for the two-population fits (bottom right). The vertical solid lines denote 1\,\re of the NSC, the dotted lines show the outer limit of the kinematic data,  the dot-dashed line in the top right panel represents the median of $R_\Upsilon$, and in the bottom right panel the median of $R_h$.}
  \label{fig:massenc1}
\end{figure*}

We show the deprojected cumulative total mass profiles with a spherical radius in Fig.~\ref{fig:massenc1} with a constant $\Upsilon$ and a free \mbh (top left panel), or with varying $\Upsilon$ and fixed \mbh (top right panel), and for comparison, we show several models from the literature. Our mass profiles are lower than the literature profiles in the region $\sim$5--20\,pc, for some, even out to 60\,pc. 
This is probably caused by the lack of data in this region in the literature. They had sparse data at best in this region (e.g. $\sim$120 stars at $r\gtrsim$6\,pc in \citealt{2014A&A...570A...2F}, $\sim$200 stars at $r\gtrsim$3.5\,pc in \citealt{2015MNRAS.447..948C}, and fewer than 140 stars at $r$\textless20\,pc in \citealt{2022MNRAS.512.1857S}). A low number of stars that may be contaminated by foreground stars can cause an overestimation of the velocity dispersion, and hence, of the total dynamical mass.
The cumulative mass of the model with a DM cusp is shown in the bottom left panel.  We also list the enclosed total mass at four different spherical deprojected radii in Table~\ref{tab:masses} and the mass for projected radii in Table~\ref{tab:masses2d}. Our resulting masses mostly agree with each other within the uncertainties.

\begin{table}
	\centering
	\caption{Enclosed total mass in units of $10^7$\,\msun at different deprojected spherical radii and for different models.}
	\label{tab:masses}
	\begin{tabular}{lcccc} 
	\hline
 Model& 5\,pc& 10\,pc& 20\,pc&30\,pc \\

 \hline
\noalign{\smallskip}  

Free \mbh &   1.18$\pm$0.03  &  1.87$\pm$0.05  &3.7$\pm$0.1  &6.8$\pm$0.2    \\
\noalign{\smallskip}
$\Upsilon(r)$ &   1.37$_{-0.12}^{+0.25}$  &  1.84$_{-0.08}^{+0.13}$  &3.6$\pm$0.2  &6.5$\pm$0.3    \\
\noalign{\smallskip}
DM cusp &   1.09$_{-0.05}^{+0.06}$  &  1.76$_{-0.07}^{+0.08}$  &3.7$\pm$0.1  &6.9$\pm{0.2}$    \\
\noalign{\smallskip}
Two-Pop&   1.21$\pm$0.02  &  1.92$\pm$0.04  &3.8$\pm$0.1  &7.0$\pm$0.2    \\
\noalign{\smallskip}  

 \hline
	\end{tabular}
    \tablefoot{ We drew 1\,000 realisations from the posterior distributions, computed their enclosed mass profile, and then took the 16th and 84th percentiles for the uncertainties, and the median was taken as the mass value. }
\end{table}

\begin{table}
	\centering
	\caption{Same as Table~\ref{tab:masses}, but for the projected radius. }
	\label{tab:masses2d}
	\begin{tabular}{lcccc} 
	\hline
 Model& 5\,pc& 10\,pc& 20\,pc&30\,pc \\

 \hline
\noalign{\smallskip}  

Free \mbh &   2.05$_{-0.06}^{+0.05}$  &  4.3$\pm$0.1  &10.7$\pm$0.4  &18.4$_{-0.7}^{+0.6}$    \\
\noalign{\smallskip}
$\Upsilon(r)$ &   2.5$_{-0.2}^{+0.5}$  &  4.2$_{-0.2}^{+0.3}$  &10.4$_{-0.5}^{+0.6}$  &17.6$_{-0.8}^{+0.9}$    \\
\noalign{\smallskip}
DM cusp &   1.78$_{-0.14}^{+0.16}$  &  3.6$_{-0.3}^{+0.4}$  &9.0$_{-0.9}^{+1.0}$  &15.4$_{-1.6}^{+1.8}$    \\
\noalign{\smallskip}
Two-Pop&   2.11$\pm$0.04  &  4.4$\pm$0.1  &11.1$\pm$0.3  &19.1$\pm$0.5    \\
\noalign{\smallskip}  

 \hline
	\end{tabular}
\end{table}

\section{Two-population dynamical models}
\label{sec:res2pop}
In this section, we describe the results we obtained when we fitted two stellar populations and an additional background component. As a chemical tracer, we used the overall metallicity \mh. 

For each population, we fitted a separate radially constant $\beta^k_z$ and $\kappa^k$. The two populations shared the same radially constant $\Upsilon$. The black hole mass was fixed to the literature value, and any DM contribution was neglected. 
The population fraction $h$ was left free to vary as a function of radius, parametrised with Eq.~\ref{eq:2val}. 
Our results are summarised in Table~\ref{tab:res2pop}.

\begin{table}
	\centering
	\caption{Results of the two-population chemo-dynamical models with an additional background component. }
	\label{tab:res2pop}
	\begin{tabular}{lccc} 
	\hline
Parameter & Unit& \mh-rich     &\mh-poor \\
   & &$k=1$ &$k=2$ \\ 
 \hline
\noalign{\smallskip}  
$\epsilon$ &[\%]     & \multicolumn{2}{c}{1.71$_{-0.15}^{+0.16}$}  \\
\noalign{\smallskip}
$\beta_z^k$ &    ...   & -0.06$_{-0.04}^{+0.03}$  & 0.64$_{-0.08}^{+0.06}$ \\
\noalign{\smallskip}
$\kappa^k$ &     ...  &-1.07$\pm$0.04   &-0.59$_{-0.18}^{+0.16}$  \\
\noalign{\smallskip}
$\Upsilon$ &    ...   &  \multicolumn{2}{c}{0.75$\pm$0.02}   \\
\noalign{\smallskip}
$h^k_0$   &      ...  & 0.98$\pm$0.01  &0.02$\pm$0.01  \\
\noalign{\smallskip}
$h^k_\infty$   &      ...  & 0.91$\pm$0.01  &0.09$\pm$0.01  \\
\noalign{\smallskip}
$R_h$    &      [arcsec]  & \multicolumn{2}{c}{45$^{+32}_{-16}$}  \\
\noalign{\smallskip}
$Z_\text{0}^k$ &[dex]    & 0.34$\pm$0.01  & -0.78$\pm0.05$\\
\noalign{\smallskip}
$\sigma_{\text{Z}}^k$ &[dex] & 0.30$\pm$0.01  & 0.27$\pm$0.03\\

\noalign{\smallskip}  

 \hline
	\end{tabular}
    \tablefoot{ The listed values are the median values of the posterior distributions, and the uncertainties are the 16th and 84th percentiles.}
\end{table}
\begin{figure}[h!]
\centering
 \includegraphics[width=0.9\columnwidth]{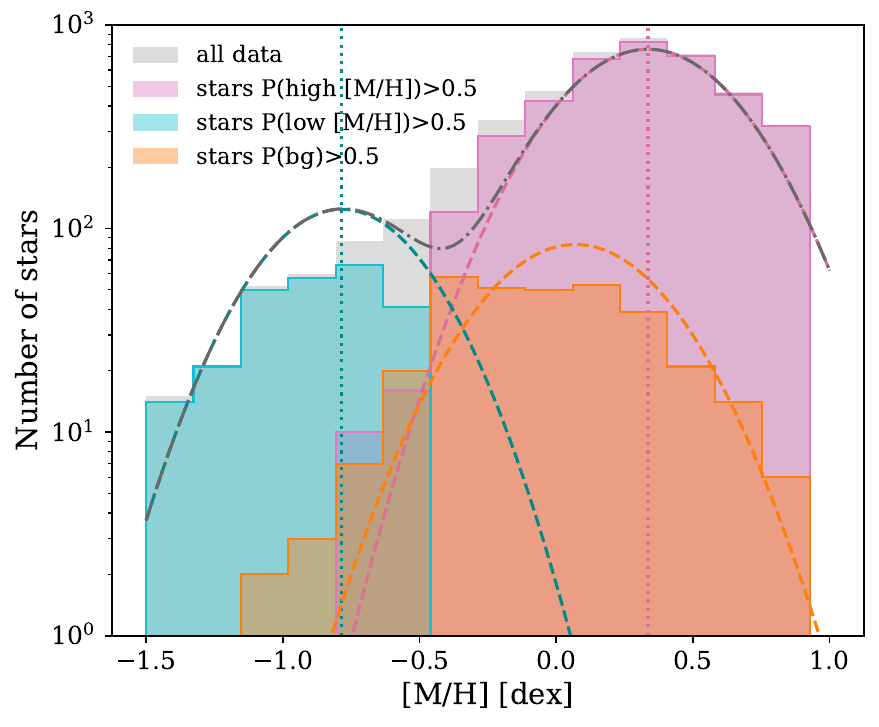}
  \caption{\mh histogram. The grey histogram denotes all stars, the vertical dotted lines show the 50th percentile of $Z^k$, and the dashed lines represent the resulting distributions of $Z^k$ and $\sigma_Z^k$.  Pink denotes the high-\mh population, cyan shows the low-\mh population, and orange shows the background population.
  }
  \label{fig:histmh}
\end{figure}

The fraction of high-\mh stars is high, more than $\sim$91\%, which corresponds to $\sim$3\,950 stars with a probability higher than 50\% of belonging to this group. Only $\sim$250 stars have a probability higher than 50\% of belonging to the low-\mh population, and $\sim$330 stars may belong to the background. About 70 stars cannot be attributed to either group and have P$^k_i$\textless50\%.

\begin{figure*}
 \centering
 \includegraphics[width=18cm]{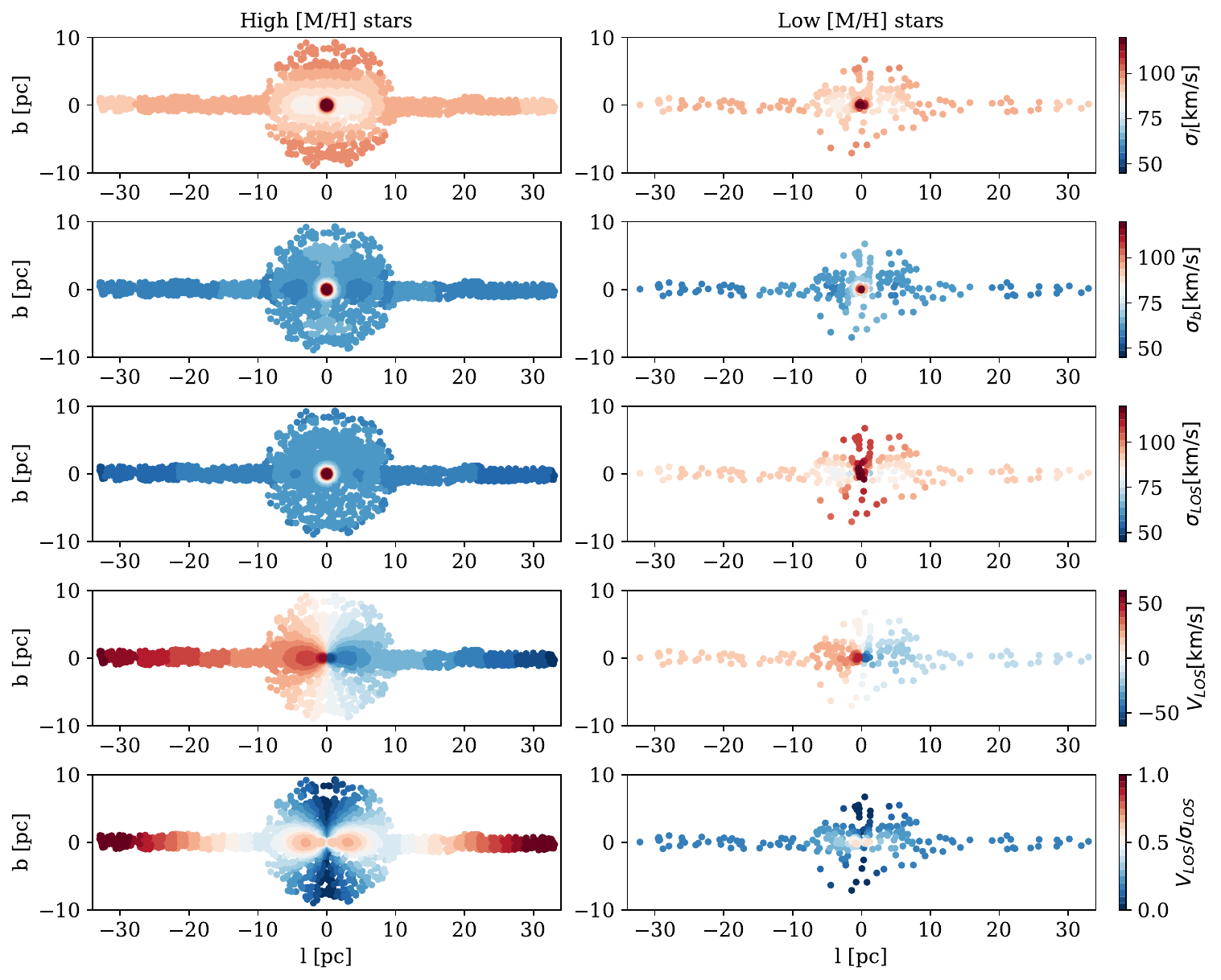}
  \caption{Spatial distribution of stars, colour-coded from top to bottom by the velocity dispersion along $l$, along $b$, along the line of sight, \vlos, \vlos/\slos, for the 50th percentile two-population model, with high-\mh stars (left), and with low-\mh stars (right). 
  }
  \label{fig:modelmh}
\end{figure*}

The two populations have a similar \mh width $\sigma_{Z}^k$. The \mh histogram is shown in Fig.~\ref{fig:histmh}. Both populations co-rotate, but the high-\mh population has a higher ratio of ordered-to-disordered motion (see also the fourth row of Fig.~\ref{fig:modelmh}). Further, the velocity anisotropy is significantly different: The anisotropy is mildly tangential for the high-\mh population (slightly negative $\beta_z^1$), and it is significantly radial for the low-\mh population (positive $\beta_z^2$). 

\begin{figure}
\centering
 \includegraphics[width=0.95\columnwidth]{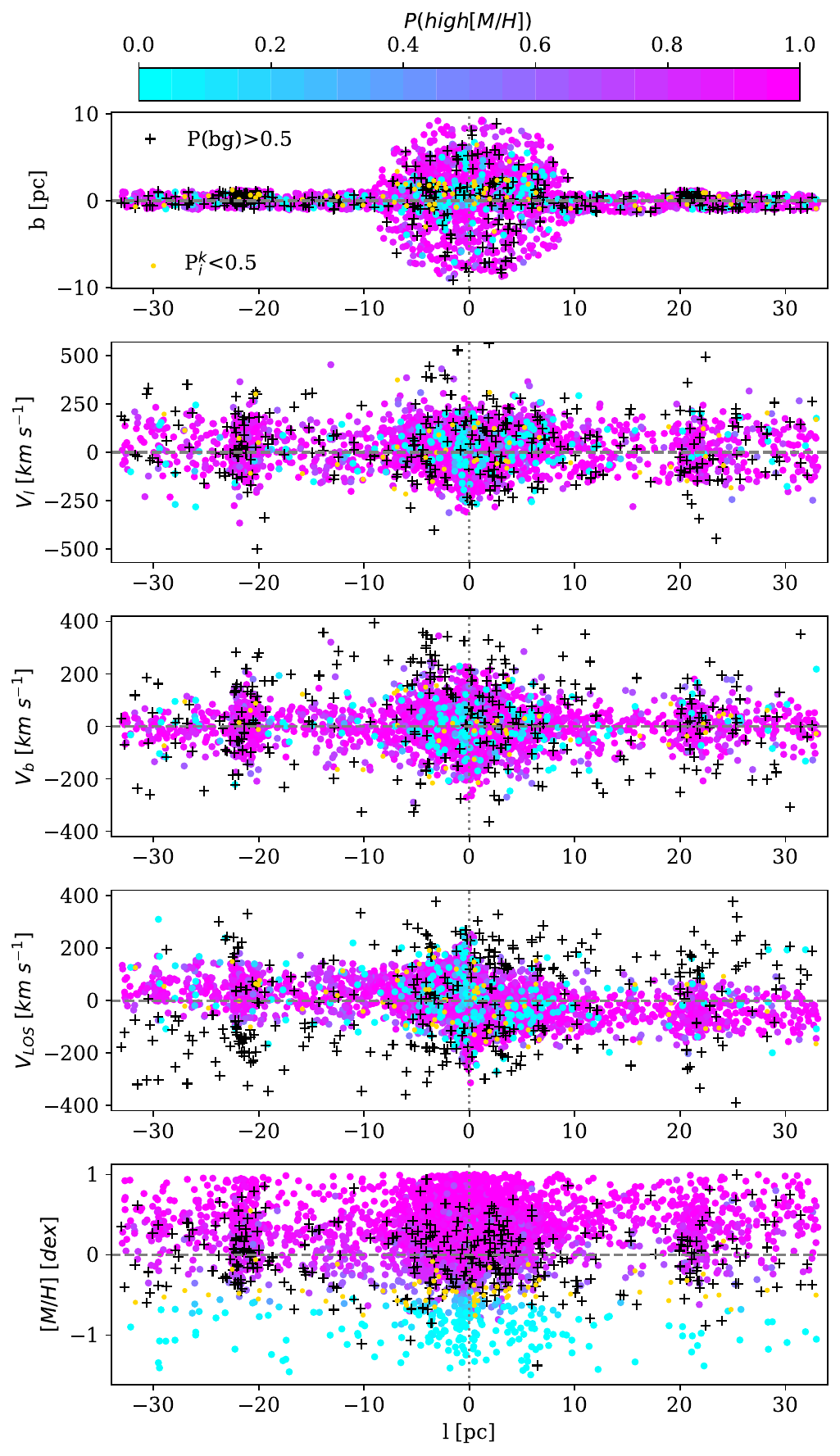}
  \caption{
  Map of the stellar positions (top), position-velocity plots along the Galactic longitude $l$ for the PM along $l$ (second panel), along $b$ (third panel), along the line of sight (fourth panel), and  along the stellar metallicity \mh. The colour-coding is from the 50th percentile realisation of the two-population model and is shown by the colour bar on the top. Pink denotes a high probability to be a high \mh star, cyan shows a low \mh star, and the black plus shows a background star. The orange circles denote stars with a population membership probability P$^k_i$\textless0.5.
    }
 \label{fig:bgcontr2mh}
\end{figure}

$\Upsilon$ is slightly higher than for the one-population models. The background fraction $\epsilon$ is significantly lower, only 1.71\%. Some of the stars that are attributed to the background in the one-component models, including those on apparently counter-rotating orbits, are attributed to the radial anisotropic low-\mh population in the two-population models (compare Figs.~\ref{fig:bgcontr} and \ref{fig:bgcontr2mh}).
The high-\mh population is dominant at all radii, but it is slightly more dominant in the centre, with $h_0^1$=0.98, after which it decreases to $h_\infty^1$=0.91. The transition radius of $h$ lies within the NSC, with $R_h\approx$1.8\,pc.

The high-\mh population resembles the one-population model in terms of the strong rotation signal and the shape of the \vlos/\slos map. The highest value of \vlos/\slos is again at the large $\lvert l \lvert$, but it is now lower and only reaches $\sim$1.1 and not 1.25, as it does in the one-population fit. On the other hand, the low-\mh population has a higher \slos ($\sim$80--90\,\kms) than the high-\mh stars (50--60\,\kms; see Fig.~\ref{fig:modelmh}, third row), except in the innermost parsec region.

\section{Discussion}
\label{sec:discussion}

Our discrete axisymmetric Jeans models constrained the central Galactic gravitational potential and the velocity structure of the stars very well. 
Our enclosed mass profile in the radial range of 5--30\,pc is significantly lower than in previous studies, up to $\sim$3.5$\times 10^7$\,\msun below the \cite{2020MNRAS.499....7S} profile at $\sim$30\,pc (using the two-population mass profile as reference), although the largest discrepancy in terms of percent is at 15\,pc, where the enclosed mass we found is lower by almost 75\%.  The agreement is better with a discrepancy of only $\sim$1.5$\times 10^7$\,\msun (at $\sim$17\,pc) with the \cite{2022MNRAS.512.1857S} mass profile, and the mass is $\lesssim$60\% lower (highest discrepancy at 12\,pc) when we added a black hole mass of $4.3\times 10^6$\,\msun to their model. The triaxial orbit-based model of \cite{2017MNRAS.466.4040F} agrees better with our results, with a discrepancy of $\sim$$7\times 10^6$\,\msun within the range of the data at most, and $\lesssim$40\% (highest discrepancy at 13\,pc). Beyond the extent of our data, the different models diverge further in terms of cumulative mass, but their relative differences are $\lesssim$25\% at 60\,pc. 
As noted above, these studies were based on a smaller sample size than ours in the range of 5--30\,pc. Their samples might also be contaminated with high-velocity stars, which are attributed to the background component that represents the Milky Way bar in our model. Both factors can bias the velocity dispersion to higher values, which then results in an overestimation of the enclosed mass.

Our result for \mbh perfectly fits the value from stellar orbit fitting, which is remarkable considering the discrepant results of other stellar dynamical models. For example, \cite{2014A&A...570A...2F} found (1.7$^{+1.4}_{-1.1})\times 10^6$\,\msun, \cite{2015MNRAS.447..948C} (3.86$\pm$0.14)$\times 10^6$\,\msun, \cite{2017MNRAS.466.4040F} (3.0$^{+1.0}_{-1.3})\times 10^6$\,\msun, and \cite{2019MNRAS.484.1166M} (3.76$\pm$0.22)$\times 10^6$\,\msun. Our mass-to-light conversion factor $\Upsilon$ also perfectly agrees with expectations for the 4.5\,\micron band (to which we scaled our number density distribution) from stellar population studies \citep{2014ApJ...788..144M,2014ApJ...797...55N,2018MNRAS.473..776K}. 

Our fits with DM suggest that a core profile contributes a few percent to the total mass in the inner $\sim$30\,pc at best. Simulations excluded a core profile \citep{2025arXiv250114868H} and favoured a steeper inner profile ($\gamma$=0.5 to 1.3), however, and our cusp profile is in this range. With a cusp profile, the DM contribution in our models is higher and reaches up to 6--25\% at 33\,pc. 
The DM volume density of simulations is not resolved in the region of our data, and \cite{2025arXiv250114868H} only showed it at $r$$\gtrsim$500\,pc. We extrapolated our cusp DM profile and found that our DM volume density is higher by about a factor of ten than that of \cite{2025arXiv250114868H}, and it is also higher than the DM density of the dynamical bar models by \cite{2017MNRAS.465.1621P}. Our value of $\rho_S$ has a firm upper limit, but a broader tail to lower values. This suggests that our DM contribution may be overestimated. Our other model parameters are robust and barely affected by the inclusion or exclusion of a DM component. This means that it was justified to neglect the DM contribution in the two-population models. 

The velocity anisotropy $\beta_z$ and $\kappa$ vary little as a function of radius. Most of the stars are attributed to a high-\mh population, which has very mild tangential anisotropy. Moreover, the models reported by  \cite{2022MNRAS.512.1857S} indicated a tangential anisotropy in the inner 30\,pc of the NSD.

The high-\mh population has strong rotational support, and it is close to an isotropic rotator because $\beta_z^1=-0.06$ is close to zero and $\kappa^1=-1.07$ is close to $-1$. For an oblate isotropic rotator,  we can use the minor-to-major axis ratio $q$ to estimate \vlos/\slos. For the NSC, we have $q$=0.7, and thus \vlos/\slos=$\sqrt{(1-q)/q}$=0.65, for the NSD, $q$=0.35, and thus \vlos/\slos=1.36. 
The Jeans model predicts that \vlos/\slos has a maximum within 1\,\re of the NSC at $\sim$0.65 (see also Fig.~\ref{fig:modelmh}, bottom left panel). 
This agrees with extragalactic NSCs in late-type galaxies, which have integrated values of (\vlos/\slos)$_e$ (i.e. within 1\,\re) of 0.2--0.55 \citep{2021ApJ...921....8P}. These values are  lower than our maximum of \vlos/\slos, as \cite{2021ApJ...921....8P} reported the flux-weighted integrated \vlos/\slos within 1\,\re, and these NSCs are not seen perfectly edge-on, as is the case for the Milky Way.  
Farther out, after a drop to \vlos/\slos$\sim$0.43 at $l\sim$9.5\,pc, we found that \vlos/\slos predicted by the model increases even further to more than $\sim$1. There is likely a maximum of \vlos/\slos, located in the NSD, but this is beyond the range of our data. 
In extragalactic NSDs, observations have detected maxima of \vlos/\slos, with values in the range of 1--3 and  at radii of 200--1\,000\,pc  \citep{2020A&A...643A..14G}, which agrees with hydrodynamical simulations of a bar-built NSD \citep{2014MNRAS.445.3352C}. Our best-fit model is consistent with the observations of extragalactic NSDs, and the high \vlos/\slos  of the high-\mh population supports the scenario that these stars formed in situ after gas inflow from the Galactic disc.  
To determine where the \vlos/\slos of the NSD has its maximum and what the maximum is, we need to model a larger region of the NSD. Kinematic data are available in the literature \citep{2021A&A...649A..83F}, and we plan to do this in the future. 

Our low-\mh population, on the other hand, is radially anisotropic and has weaker rotational support (\vlos/\slos $\lesssim$0.4 at $l\gtrsim$2\,pc), with an overall mean of \vlos/\slos=0.2. The reason is twofold: The value of \vlos near the Galactic plane ($\lvert b \lvert \lesssim$0.5\,pc) is $\sim$$\pm$16\,\kms, but $\sim\pm$24--55\,\kms for the high-\mh population, whereas \slos is $\sim$90\,\kms, but 50--60\,\kms for the high-\mh population. The velocity dispersions $\sigma_l$ and $\sigma_b$ are not significantly different for the two populations. 

The fraction of background stars in the two-population models is lower than in the one-population model, and some of the stars attributed to the low-\mh population were previously assigned to the background component. The question therefore arises whether some stars of the 
low-\mh population are instead bar interlopers, with less extreme velocities and rather low \mh. 
While this may be the case for some stars, we found that the distributions of the low-\mh  (i.e. stars with $P^{2}\geq 0.5$) and the background populations (i.e. stars with $P^{bg}> 0.5$) significantly different. We performed $k$-sample Anderson-Darling tests \citep{Scholz01091987} for $V_l$, $V_b$, \vlos, \mh, and the distance $r$ to \sgra for the low-\mh and the background populations. In each of these tests, we obtained low $p$ values, and the null hypothesis that the low-\mh and the background population samples come from the same distribution can be rejected with a confidence of 99.9\%. We conclude that it is unlikely that the entire low-\mh population can be attributed to the background bar component. 

One possible explanation for the low-\mh population is an external origin. Dynamical friction can move massive star clusters towards the Galactic centre, where the stars mix with the in situ stellar population \citep[e.g.][]{1975ApJ...196..407T,1993ApJ...415..616C,2012ApJ...750..111A,2020ApJ...901L..29A,2024MNRAS.529.4104V}. \cite{2017MNRAS.464.3720T} showed with N-body simulations that even an NSC built entirely from 12 consecutively infalling star clusters coming from random orbital directions can have rotational support (\vlos/\slos=0.1--0.7). Milky Way globular clusters on eccentric orbits may also approach the Galactic centre and deposit some of their stars. \cite{2023A&A...674A..70I} found that some of the known globular clusters of the Milky Way have a minimum distance to the Galactic centre of several dozen parsec. 
The ten clusters with the highest probability of such an encounter in \cite{2023A&A...674A..70I} include several with [Fe/H] in the range of --1.5 to --1.0\,dex \citep{2009A&A...508..695C}, which is close to our low-\mh sample ($-1.5$\,dex\textless[M/H]\textless$-0.5$\,dex).  Their contribution to our data set may be low. Former already-dissolved clusters may also have deposited stars in the past. 
An external origin might explain why the low-\mh population has different values for $\kappa$ and $\beta_z$. These star clusters may have been on radial orbits that brought them close to the Galactic centre, which explains the radial anisotropy of the low-\mh stars we measured. As the two-body relaxation time in the NSC is shorter than $\sim$3\,Gyr \citep{2020ApJ...901L..29A}, this suggests that such infalls or passages occurred in the last 3\,Gyr. To constrain the number of these infall events, more precise \mh measurements and chemical abundances are needed.

Our axisymmetric Jeans modelling approach has some limitations because it (1) assumes  an axisymmetric shape and (2) makes assumptions on the alignment of the velocity ellipsoid. Although axisymmetric models fit the data very well also for more extended NSD data \citep[e.g.,][]{2024MNRAS.530.2972S}, it is unclear whether the NSD is truly axisymmetric or if the Milky Way has an inner bar \citep{2001A&A...379L..44A,2011A&A...534L..14G}, as was found in several galaxies \citep{2019MNRAS.482..506G,2024MNRAS.528.3613E} at scales of 0.1--1\,kpc. \cite{2022MNRAS.512.1857S} suggested that the kinematic signature of a nuclear or inner bar may be seen in an asymmetry of the $V_l$ distribution. A larger statistical sample and a better understanding of the expected signature of a nuclear bar from theoretical studies are needed, however. 
The second limitation (assumptions on the alignment of the velocity ellipsoid) can be overcome by using orbit-based triaxial models \citep{2020MNRAS.496.1579Z}, but these models  are not yet available for discrete data.

\section{Conclusion}
\label{sec:con}
 We have constructed the first discrete axisymmetric chemo-dynamical Jeans models of the Galactic centre.
The chemo-dynamic data consisted of \vlos\ and \mh for 4\,600 unique stars, 3\,567 of which also had proper motions. Foreground and background stars were excluded based on \col\ colour cuts. We accounted for the remaining contribution of bar stars by including a background component in the models.  Our data extend out to $|l|$=33\,pc from \sgra\ along the Galactic longitude, into a region in which other studies only had several hundred stars. 

We constructed purely dynamical models for which only the stellar density distribution and  kinematic data were used. These models showed that the velocity anisotropy $\beta_z\approx-0.1$, the rotation parameter $\kappa\approx-1$, and the mass-to-light ratio $\Upsilon\approx0.7$ depend little on the radius and can be assumed to be constant. When we fitted the mass of \sgra, we obtained (4.35$^{+0.24}_{-0.23}\times10^6$)\,\msun, which perfectly agrees with literature values of stellar orbits. 
We tested a cored ($\gamma$=0) and a cusped ($\gamma$=1) DM profile, but the mass contribution in the NSC was only a few percent in either case, and we decided to neglect DM. 
We listed the enclosed total mass of various models for several deprojected and projected radii. 

In chemo-dynamical models, we assumed two populations $k$ of stars, each with a unique but radially constant value for $\beta_z^k$ and $\kappa^k$, and Gaussian \mh distributions with a unique centroid $Z_0^k$ and width $\sigma_Z^k$. The two populations shared the same $\Upsilon$; their relative contribution was a function of radius. We obtained a dominant ($\sim$90\%) \mh-rich population with a mild tangential anisotropy ($\beta_z^1$=--0.06) and strong rotation ($\kappa^1=-1.06$), which is consistent with being formed in situ after gas inflow from the outer Galactic disc transported in via the bar. The sub-dominant \mh-poor population has a significant radial anisotropy ($\beta_z^2$=+0.64) and is less well supported by rotation ($\kappa^2=-0.6$), which indicates a different origin.

For a better understanding of the Galactic centre mass distribution and orbital structure, it would be interesting to relax some assumptions of the axisymmetric Jeans models and use triaxial orbit-based models, for example, which make no assumptions on the intrinsic velocity distribution. 
Concerning the data, a larger sample with more precise velocity and \mh measurements and possibly even chemical abundances would allow us to obtain a more thorough analysis of the subpopulations and even to include more than two populations. This would help us to better understand the history of merger and accretion events, and thus, the formation of the nuclear region of the Milky Way.

\begin{acknowledgements}
We thank the anonymous referees for their thorough review and constructive comments that  improved the manuscript.
AFK acknowledges funding from the Austrian Science Fund (FWF) [grant DOI 10.55776/ESP542].
This research made use of NumPy \citep{harris2020array}, SciPy \citep{Virtanen_2020}, matplotlib, a Python library for publication quality graphics \citep{Hunter:2007}, Astropy, a community-developed core Python package for Astronomy \citep{2013A&A...558A..33A,2018AJ....156..123A, 2022ApJ...935..167A}.
 
 \end{acknowledgements}


\bibliographystyle{aa}
\bibliography{example}

%%%%%%%%%%%%%%%%%%%%%%%%%%%%%%%%%%%%%%%%%%%%%%%%%%

%%%%%%%%%%%%%%%%% APPENDICES %%%%%%%%%%%%%%%%%%%%%

\newpage
\onecolumn
\noindent
%%%%%%%%%%%%%%%%%%%%%%%%%%%%%%%%%%%%%%%%%%%%%%%%%%

\begin{appendix}

\section{MCMC post burn-in  distributions}
\begin{figure*}
 \centering
 \includegraphics[width=19cm]{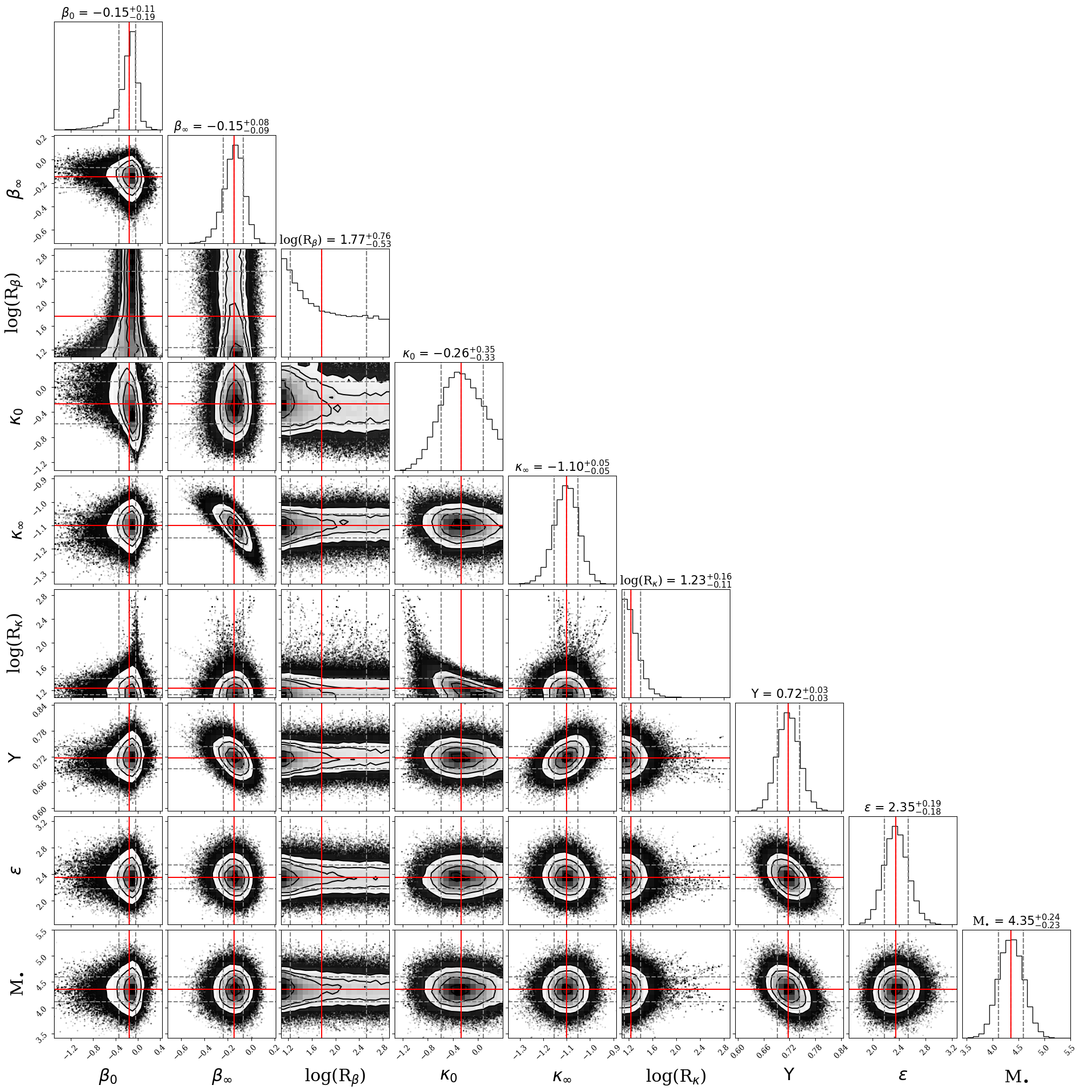}
  \caption{MCMC post burn-in  distributions for the one-population models with \mbh\ as free parameter  (Sect.~\ref{sec:fitbh}). The scatter plots show the projected two-dimensional distributions, the red lines the respective 50th percentile, and the dashed grey lines the 16th and 84th percentiles. Their values are also written on top of each column, displaying the projected one-dimensional distributions. From top to bottom and left to right, the panels show the inner anisotropy $\beta_0$, outer anisotropy $\beta_\infty$, the anisotropy transition radius $\log(R_\beta)$, inner rotation parameter $\kappa_0$, outer rotation parameter $\kappa_\infty$, the rotation transition radius $\log(R_\kappa)$, the mass-to-light ratio $\Upsilon$, fraction of background stars $\epsilon$ in percent, and the black hole mass \mbh\ in $10^6$\,\msun. }
  \label{fig:cornerbh}
\end{figure*}

\begin{figure*}
 \centering
 \includegraphics[width=19cm]{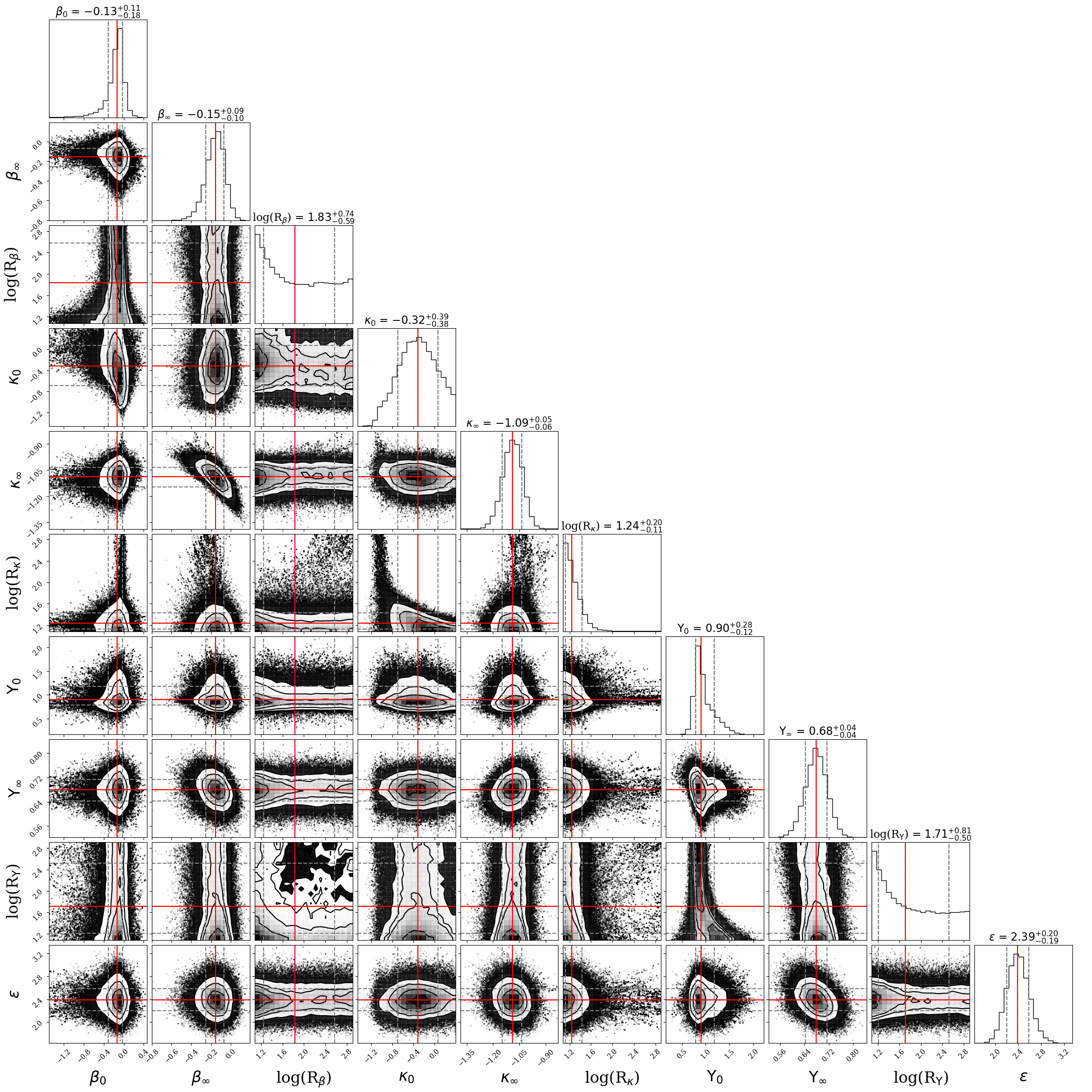}
  \caption{Same as Fig.~\ref{fig:cornerbh}, but for the one-populations models with a varying three-parameter mass-to-light ratio (Sect.~\ref{sec:fitmlr}).  From top to bottom and left to right, the panels show the inner anisotropy $\beta_0$, outer anisotropy $\beta_\infty$, the anisotropy transition radius $\log(R_\beta)$, inner rotation parameter $\kappa_0$, outer rotation parameter $\kappa_\infty$, the rotation transition radius $\log(R_\kappa)$, the inner mass-to-light ratio $\Upsilon_0$, outer mass-to-light ratio $\Upsilon_\infty$,  mass-to-light ratio transition radius $\log(R_\Upsilon)$, and the fraction of background stars $\epsilon$ in percent. The black hole mass is fixed to 4.3$\times 10^6$\,\msun, dark matter is not considered in the gravitational potential.}
  \label{fig:cornerml3}
\end{figure*}

\begin{figure*}
 \centering
\includegraphics[width=19cm]{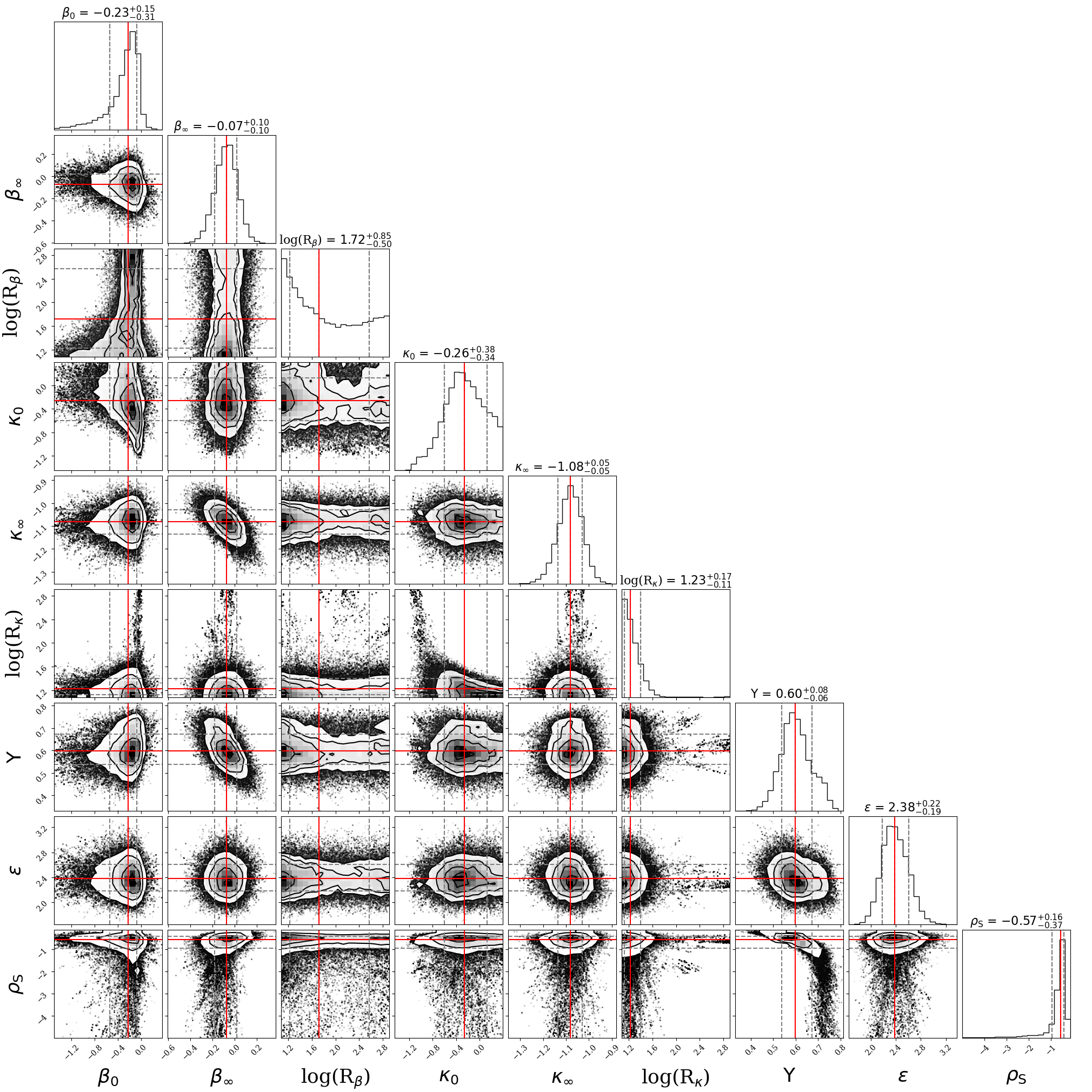}
  \caption{Same as Figs.~\ref{fig:cornerbh}--\ref{fig:cornerml3}, but for the one-population models with a cuspy dark matter profile  (Sect.~\ref{sec:fitdm}).  From top to bottom and left to right, the panels show the inner anisotropy $\beta_0$, outer anisotropy $\beta_\infty$, the anisotropy transition radius $\log(R_\beta)$, inner rotation parameter $\kappa_0$, outer rotation parameter $\kappa_\infty$, the rotation transition radius $\log(R_\kappa)$, the mass-to-light ratio $\Upsilon$, fraction of background stars $\epsilon$ in percent, and the dark matter scale density $\rho_S$. The black hole mass is fixed to 4.3$\times 10^6$\,\msun.}
  \label{fig:cornerdmcusp}
\end{figure*}

\begin{figure*}
 \centering
\includegraphics[width=19cm]{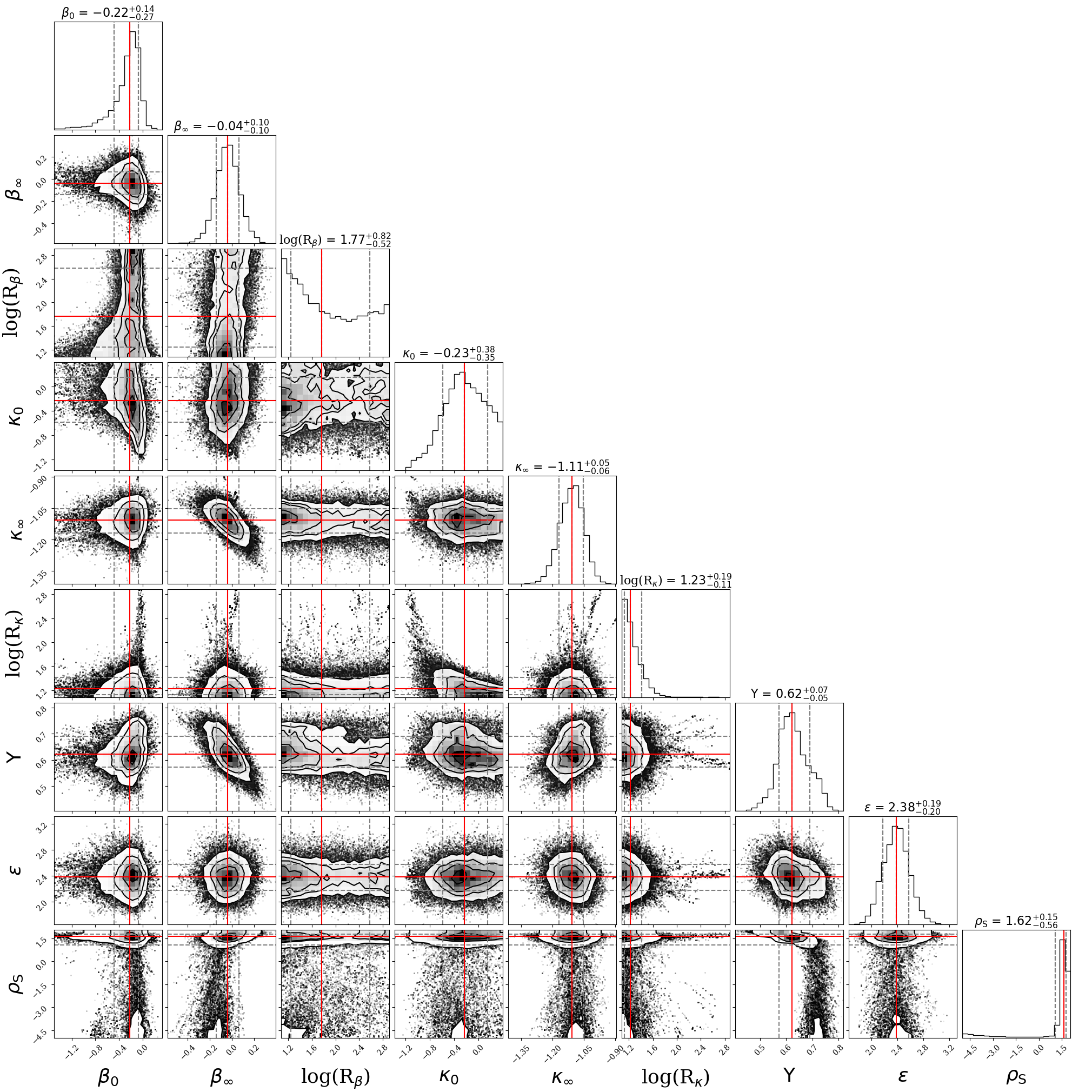}
  \caption{Same as Fig.~\ref{fig:cornerdmcusp}, but for the one-population models with a cored dark matter profile  (Sect.~\ref{sec:fitdm}).}
  \label{fig:cornerdmcore}
\end{figure*}

\begin{figure*}
 \centering
\includegraphics[width=19cm]{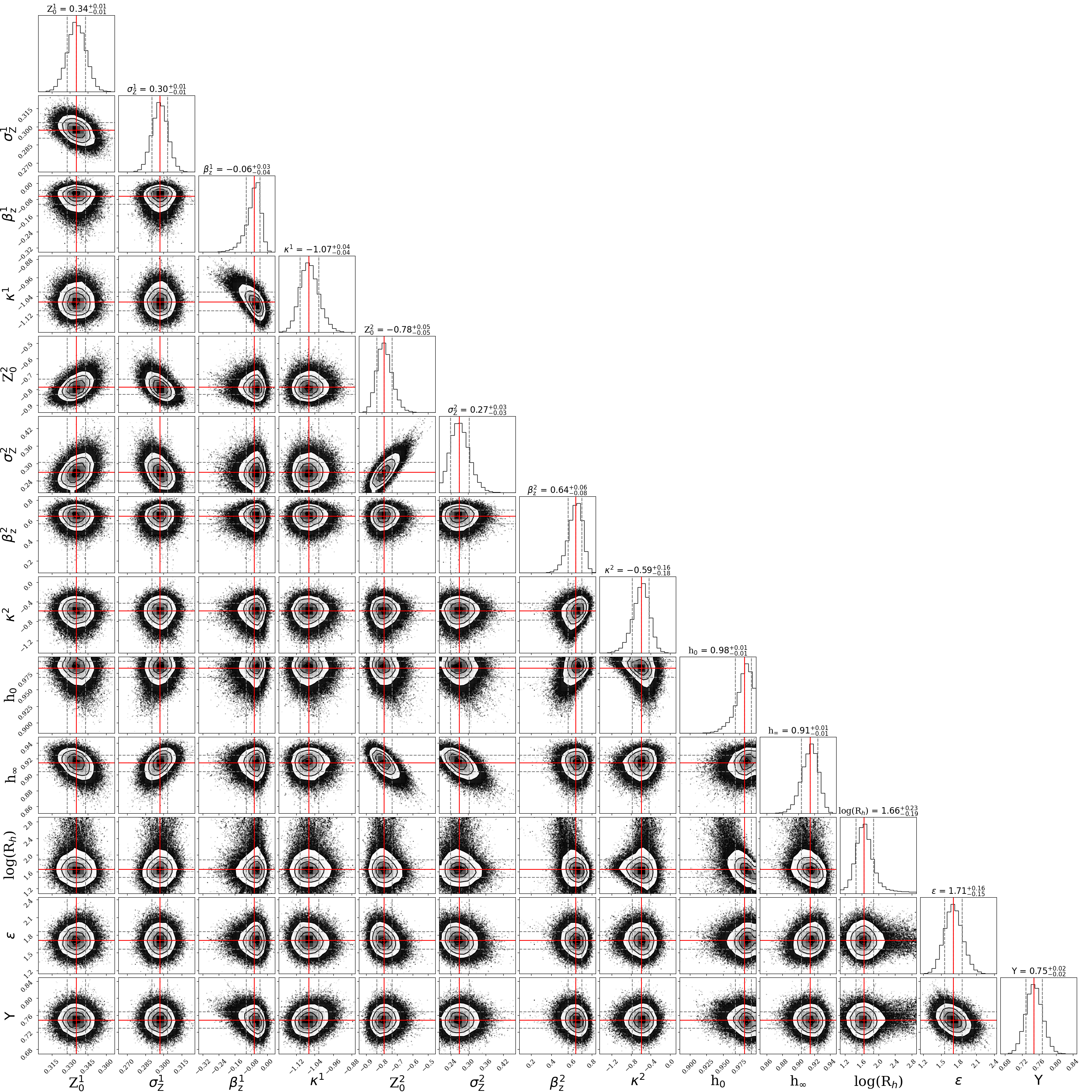}
  \caption{Same as Figs.~\ref{fig:cornerbh}-\ref{fig:cornerdmcore}, but for the two-population models (Sect.~\ref{sec:res2pop}).  From top to bottom and left to right, the panels show the mean metallicity $Z_0^1$, metallicity dispersion $\sigma_Z^1$,  anisotropy $\beta_z^1$,  rotation parameter $\kappa^1$ for the high \mh population, $Z_0^2$,  $\sigma_Z^2$,   $\beta_z^2$,  $\kappa^2$ for the low \mh population, inner  population fraction $h_0$, outer population fraction $h_\infty$, the  transition radius $\log(R_h)$,  fraction of background stars $\epsilon$ in percent, and the mass-to-light ratio $\Upsilon$. The black hole mass is fixed to 4.3$\times 10^6$\,\msun, dark matter is not considered in the gravitational potential.}
  \label{fig:cornerpop2}
\end{figure*}

In Figs. \ref{fig:cornerbh}-\ref{fig:cornerpop2} we show the post burn-in probability distributions of our various \textsc{emcee} fits.


\end{appendix}

\end{document}